\documentclass[useAMS,usenatbib]{mn2e}

\usepackage{graphicx}
\usepackage{txfonts}

\title[C method]
      {``Counterpart'' method for abundance determinations in  
H\,{\Large II} regions} 

\author[L.~S.~Pilyugin, E.~K.~Grebel, L.~Mattsson]
       {L.S.~Pilyugin$^{1,2}$,
        E.K.~Grebel$^{2}$,
        L.~Mattsson$^{3}$,  \\ 
     $^{1}$ Main Astronomical Observatory
            of National Academy of Sciences of Ukraine,
            27 Zabolotnogo str., 03680 Kiev, Ukraine \\
     $^{2}$ Astronomisches Rechen-Institut, Zentrum f\"{u}r Astronomie 
           der Universit\"{a}t Heidelberg, 
           M\"{o}nchhofstr.\ 12--14, 69120 Heidelberg, Germany \\
     $^{3}$ DARK Cosmology Centre, Niels Bohr Institute,
            University of Copenhagen, Juliane Maries Vej 30,
            DK-2100, Copenhagen \O, Denmark\\
              }

\date{Accepted 2012 May 25. Received 2012 May 09; in original form 2012 March 01}

\begin{document}

\maketitle

\begin{abstract}
We suggest a new way of the determining abundances and electron temperatures in H\,{\sc ii} 
regions from strong emission lines. Our approach is based on the standard assumption
that  H\,{\sc ii} regions with similar intensities of strong emission lines have similar physical 
properties and abundances. A ``counterpart'' for a studied H\,{\sc ii} region may be chosen 
among H\,{\sc ii} regions with well-measured abundances (reference H\,{\sc ii} regions)   
by comparison of carefully chosen combinations of strong line intensities. Then the abundances in the 
investigated  H\,{\sc ii} region can be assumed to be the same as those in its counterpart.  
In other words, we suggest to  determine the abundances in H\,{\sc ii} regions ``by precedent''. 
To get more reliable abundances for the considered H\,{\sc ii} region, 
a number of reference H\,{\sc ii} regions is selected and then the abundances in the target
H\,{\sc ii} region are estimated through extra-/interpolation. 
We will refer to this method of abundance determination as the counterpart method or, for brevity, 
the $C$ method. We define a sample of reference H\,{\sc ii} regions 
and verify the validity of the $C$ method. We find that this method produces reliable abundances. 
Finally, the $C$ method is used to obtain the radial abundance distributions in 
the extended discs of the spiral galaxies M~83, NGC~4625 and NGC~628. 
\end{abstract}



\section{Introduction}

Metallicities play a key role in many studies of galaxies.
While absorption line indices are widely used to derive the metallicities
of older stellar populations, 
gas-phase oxygen abundances are the best means to estimate 
the present-day metallicities. Since 
emission lines in the spectra of H\,{\sc ii} regions are easily 
measurable across a wide range of extragalactic distances, 
they are generally considered the most powerful 
indicators of the present-day chemical composition of star-forming galaxies. 
The spectra of a large number of individual H\,{\sc ii} regions in nearby 
spiral and irregular galaxies have now been obtained 
\citep[see][among many others]{McCall1985ApJS57,
Zaritsky1994ApJ420,vanZee1998AJ116,vanZee2006ApJ636,
Izotov1997ApJS108,Izotov1998ApJ500,Izotov2004ApJ602,Kehrig2004AJ128,
Bresolin1999ApJ540,Bresolin2005AA441,Bresolin2009ApJ695,Bresolin2009ApJ700,
LopezSanchez2009AA508,Guseva2011AA529}. 
These spectroscopic measurements provide the basis for investigations of metallicity 
properties of galaxies
such as radial abundance gradients, mean metallicities, etc., 
\citep[][among others]{Vila1992MNRAS259,Zaritsky1994ApJ420,
vanZee1998AJ116,Pilyugin2004AA425}. 

The H\,{\sc ii} regions ionised by stars (or star clusters) form a well-defined 
fundamental sequence in different emission-line diagrams. 
The existence of such a fundamental sequence provides the basis of 
various investigations of extragalactic H\,{\sc ii} regions.
In particular, \citet{baldwin1981} suggested that the position of an 
object in some well-chosen emission-line diagrams can be used to separate 
H\,{\sc ii} regions ionised by stars from other types of emission-line objects. 
This idea has found general acceptance and is widely used. Thus, the 
[O\,{\sc iii}]$\lambda$5007/H${\beta}$ vs.\ [N\,{\sc ii}]$\lambda$6584/H${\alpha}$ 
diagram is often used to distinguish between  
H\,{\sc ii} regions and active galactic nuclei (AGNs).
However, the exact location of the dividing line between H\,{\sc ii} regions 
and AGNs is still controversial  
\citep{kewley2001,kauffmann2003,Stasinska2006MNRAS371}. 
It has been argued that the binary classification scheme for 
emission line galaxies (subdividing into star-forming galaxies and AGNs) 
is oversimplified and a revised classification scheme involving more 
classes should be considered \citep{Stasinska2008MNRAS391,CidFernandes2011MNRAS413}.

\citet{Pagel1979MNRAS189} and \citet{alloin1979} suggested that the positions 
of H\,{\sc ii} regions in some emission-line diagrams can be calibrated in 
terms of their oxygen abundances. This approach to abundance determination 
in H\,{\sc ii} regions, usually referred to as the ``strong-line method'' or ``strong-line calibrations'' 
has been widely adopted. 
Numerous relations have been suggested to convert 
metallicity-sensitive emission-line ratios into metallicity or 
temperature estimates \citep[][among many others]{Dopita1986ApJ307,McGaugh1991ApJ380,
Zaritsky1994ApJ420,Pilyugin2000AA362,Pilyugin2001AA369,Kewley2002ApJS142,
Pilyugin2005ApJ631,Pettini2004MNRAS348,Tremonti2004ApJ613,Stasinska2006AA454}. 

It should be stressed that strong-line calibrations for oxygen abundances do not
form a uniform "family". Basically, there are two types. The calibrations of the first type are 
the empirical calibrations, established on the basis of H\,{\sc ii} regions in which the
oxygen abundances are determined through the $T_e$  method. 
The calibrations of the second type are
the theoretical (or model) calibrations, established on the basis of grids
of photoionisation models of H\,{\sc ii} regions. 
Among published strong-line calibrations there exist large systematic discrepancies, 
in the sense that theoretical calibrations generally produce oxygen abundances that are by
factors of 1.5 -- 5 higher than those derived using empirical calibrations
\citep[c.f.][]{Kennicutt2003ApJ591,Pilyugin2003AA399,Yin2007AA462,Kewley2008ApJ681,
Bresolin2009ApJ700,Moustakas2010ApJS190,Lopezsanchez2010AA517}. 
Thus, at the present time there exists no absolute scale for metallicities in H\,{\sc ii} regions.

The empirical metallicity scale has advantages as compared 
to the theoretical (model) metallicity scales. The empirical metallicity scale 
is well defined in terms of the abundances in H\,{\sc ii} regions derived through the  $T_e$ method,
i.e., in that sense the empirical metallicity scale is absolute. The abundances estimated via 
different empirical calibrations are compatible with each other and with the  $T_e$-based 
abundances as well. Contrary to the consistency among empirical calibrations, there are as many 
theoretical (model) metallicity scales as there are sets of H\,{\sc ii} region models.  In other words,
the abundances derived using different theoretical calibrations are usually not in agreement with each other. 
However, the  validity of the $T_e$ method (and, as a consequence, the validity of the 
empirical metallicity scale in H\,{\sc ii} regions) has for long been questioned 
\citep[][and references therein]{Peimbert1967ApJ150,Stasinska2005AA434,PenaGuerrero2012ApJ746}.  
But there is also evidence that the classic $T_e$ method 
provides realistic oxygen abundances of H\,{\sc ii} regions 
\citep{Pilyugin2003AA399,Pilyugin2006MNRAS367,Williams2008ApJ677,Bresolin2009ApJ700,Rodriguez2010ApJ708}. 
It is also noteworthy that oxygen abundances in Galactic H\,{\sc ii} regions derived using the direct $T_e$ method as well as empirical calibrations 
agree with stellar oxygen abundances \citep[see, e.g., Fig. 7 in][]{Mattsson2010}
determined in Cepheids \citep{Andrievsky2002a,Andrievsky2002b,Andrievsky2002c,Andrievsky2004} 
as well as the new solar oxygen abundance \citep{Asplund2009}.
Moreover, if the empirical metallicity scale should be corrected \citep[see, e.g.][]{PenaGuerrero2012ApJ746},
the abundances derived using the $T_e$ method and different empirical calibrations 
should be corrected accordingly. 
Thus, the empirical metallicity scale is likely the preferable metallicity scale at present.

However, all calibrations (empirical as well as theoretical) encounter problems. 
These calibrations are usually based on the oxygen [O\,{\sc ii}]$\lambda$3727+$\lambda$3729, 
[O\,{\sc iii}]$\lambda$5007 and/or nitrogen [N\,{\sc ii}]$\lambda$6584 lines 
[with a few exceptions, e.g., \citet{Stasinska2006AA454}]. 
It is well known that the relation between the oxygen abundance and the strong 
oxygen-line intensities is double-valued, with two distinct parts,
traditionally known as the upper (12 + log(O/H) $\ga$ 8.25) and lower (12 + log(O/H) $\la$ 8.0)  
branches of the R$_{23}$ -- O/H diagram. Moreover, the strong oxygen-line intensities 
are not a good indicator of the oxygen abundance in the transition zone between the upper 
and lower branches. 

Furthermore, it is well known that there is no one-to-one correspondence between oxygen 
and nitrogen abundances. A prominent feature of the N/O vs. O/H diagram is that 
the N/O abundance ratio shows a large scatter at a 
fixed value of the O/H abundance ratio, larger than can be explained by  
observational uncertainties 
\citep{Henry2000ApJ541,Pilyugin2003AA397,Lopezsanchez2010AA517}. 
The N/O -- O/H relation shows also a clear bend: while at low 
metallicities (12 +log(O/H) $\la$ 8.0) the N/O abundance ratio is, on average, 
constant, the N/O ratio increases with O/H at high metallicities 
(12 +log(O/H) $\ga$ 8.0). 

The properties mentioned above prevent the construction of a calibration that 
works over the whole range of metallicities shown by H\,{\sc ii} regions. 
Thus, one has to construct separate calibrations for different metallicity intervals, 
i.e., there are no calibration relations that work sufficiently well over the whole 
range of observed metallicities. 
Here, also another problem arises --  one has to know {\it a priori}
in which metallicity interval (or on which of the two branches) the H\,{\sc ii} region
is located.

It should be emphasised that each existing calibration is based on the assumption
that H\,{\sc ii} regions with similar strong-line intensities have similar abundances. 
A simple, more direct method for abundance determination follows from that assumption as well. 
If there were (and fortunately there is indeed) a suitable sample of reference H\,{\sc ii} regions with well-measured 
electron temperatures and abundances, then one can choose among those 
reference H\,{\sc ii} regions the ones that have the smallest difference in strong line intensities  
compared to the studied H\,{\sc ii} region, i.e., one can find a corresponding, ``counterpart'' H\,{\sc ii} region. 
Then the oxygen and nitrogen abundances and electron temperatures in the 
investigated  H\,{\sc ii} region can be assumed to be the same as in its counterpart. 
In other words, we suggest that the abundances in the target H\,{\sc ii} region can be determined ``by precedent''. 
To obtain more reliable abundances, one may select several reference H\,{\sc ii} regions 
(counterparts) and then estimate the abundance in the target H\,{\sc ii} region through 
extrapolation or interpolation. We will refer to this method as the ``counterpart method'' 
or, for brevity, as the $C$ method.

The main goal of the present study is to select a sample of H\,{\sc ii} regions with  
well-measured electron temperatures and abundances, i.e., to obtain a sample of 
reference H\,{\sc ii} regions. We have carried out an extensive search of the literature 
to compile a list of spectra of H\,{\sc ii} regions in irregular and spiral galaxies with measured electron temperatures.  
The sample of reference H\,{\sc ii} regions and the $C$ method are discussed in Section 2. 
In Section 3, the $C$ method is used to obtain abundance gradients in the extended 
discs of three spiral galaxies. Section 4 presents the conclusions.

Throughout the paper, we will use the following standard notations for the line 
intensities: \\ 
$R$  = $I_{\rm [O\,III] \lambda 4363} /I_{{\rm H}\beta }$,  \\
$R_2$  = $I_{\rm [O\,II] \lambda 3727+ \lambda 3729} /I_{{\rm H}\beta }$,  \\
$N_2$  = $I_{\rm [N\,II] \lambda 6548+ \lambda 6584} /I_{{\rm H}\beta }$,  \\
$S_2$  = $I_{\rm [S\,II] \lambda 6717+ \lambda 6731} /I_{{\rm H}\beta }$,  \\
$R_3$  = $I_{{\rm [O\,III]} \lambda 4959+ \lambda 5007} /I_{{\rm H}\beta }$.  \\
$S_3$  = $I_{{\rm [S\,III]} \lambda 9068+ \lambda 9532} /I_{{\rm H}\beta }$.  \\
The electron temperatures will be given in units of 10$^4$K.

\section{A sample of reference H\,{\sc ii} regions}

\subsection{Observational data: line intensities}

We have carried out an extensive search of the literature and compiled a sample of H\,{\sc ii} 
regions with abundances determined with the $T_e$ method. This sample is the basis for our study.
We have searched for spectra of H\,{\sc ii} regions in irregular and 
spiral galaxies, with the requirement that they include the 
[O\,{\sc ii}]$\lambda$3727+$\lambda$3729, 
[O\,{\sc iii}]$\lambda$5007,
[N\,{\sc ii}]$\lambda$6584, 
[S\,{\sc ii}]$\lambda$6717+$\lambda$6731 lines  
and a detected auroral line of, at least, 
one ion, i.e., [O\,{\sc iii}], [N\,{\sc ii}], [S\,{\sc iii}]. The electron temperature $t_2$ 
in H\,{\sc ii} regions can also be estimated from the  
[O\,{\sc ii}]$\lambda$$\lambda$3727,3729/[O\,{\sc ii}]$\lambda$$\lambda$7320,7330 ratio. 
However, the difference between the  $t_2$ temperature derived from the 
[O\,{\sc ii}]$\lambda$$\lambda$3727,3729/[O\,{\sc ii}]$\lambda$$\lambda$7320,7330 ratio
and the one estimated from the commonly used $t_2$ -- $t_3$ relation can be large 
(see the discussion in \citet{Kennicutt2003ApJ591} and references therein), making the 
oxygen abundances so derived very uncertain. Therefore, H\,{\sc ii} regions with 
a temperature derived from the 
[O\,{\sc ii}]$\lambda$$\lambda$3727,3729/[O\,{\sc ii}]$\lambda$$\lambda$7320,7330 ratio
are not considered in the present study.
While we have tried to include as many 
sources as possible, we do not claim our search to be exhaustive. 

In recent years, the number of available spectra of emission-line nebulae has increased 
dramatically due to several large spectroscopic surveys such as 
the Sloan Digital Sky Survey (SDSS) \citep{York2000AJ120}. The auroral lines are measurable in a 
relatively large number of SDSS galaxies \citep{Kniazev2004ApJS153,Izotov2006AA448}, 
which provides the possibility to obtain $T_e$-based abundances for SDSS galaxies. 
However, the SDSS objects cannot be used as reference H\,{\sc ii} regions
for two reasons. 
First, the wavelength range of the SDSS spectra is 3800 -- 9300 \AA\ so that
for nearby galaxies with redshift $z \la 0.02$ the 
[O\,{\sc ii}]$\lambda$3727+$\lambda$3729 emission line is outside of that range. 
The lack of this line prevents us from 
using SDSS spectra of nearby galaxies in our study. 
Second, the SDSS galaxy spectra span a large range of redshifts. 
There is thus an aperture-redshift effect in SDSS spectra since these spectra are 
obtained with $3''$-diameter fibers. At a redshift of $z =0.05$ 
the projected aperture diameter is $\sim$ 3 kpc, while it is  $\sim$ 15 kpc
at a redshift of $z=0.25$. This means that, at large redshifts, 
 SDSS spectra are closer to global spectra of whole galaxies, i.e., to spectra of  
composite nebulae including multiple star clusters, 
rather than to spectra of individual H\,{\sc ii} regions.
It has been argued that the $T_e$ method can 
result in an underestimated oxygen abundance in the SDSS objects 
if H\,{\sc ii} regions with different physical properties contribute 
to the global spectrum of composite nebulae \citep{Pilyugin2012MNRAS000}. 
This effect is somewhat similar 
to the small-scale temperature fluctuations in H\,{\sc ii} regions discussed by 
\citet{Peimbert1967ApJ150}.   

High-precision spectroscopy, including the auroral lines  
[O\,{\sc iii}]$\lambda$4363 and [N\,{\sc ii}]$\lambda$5755 for  
a number of H\,{\sc ii} regions in our Galaxy 
\citep[see][among others]{Esteban2004MN355,GarciaRojas2004ApJS153,GarciaRojas2005MN362,GarciaRojas2006MN368,GarciaRojas2007ApJ670}
and in the Large and Small Magellanic Clouds 
\citep[e.g.][]{Peimbert2003ApJ584,Tsamis2003MN338,Peimbert2005ApJ634,PenaGuerrero2012ApJ746} 
can be found in the literature.   
However, only a small part of the H\,{\sc ii} regions is measured in these cases and 
therefore the obtained line intensities are usually not representative for 
the whole nebula. For this reason, these spectroscopic measurements were not 
included in our list.

Thus, for each listed spectrum, we record the measured values of 
[O\,{\sc ii}]$\lambda$3727+$\lambda$3729, 
[O\,{\sc iii}]$\lambda$4363, 
[O\,{\sc iii}]$\lambda$5007,
[N\,{\sc ii}]$\lambda$5755, 
[S\,{\sc iii}]$\lambda$6312,
[N\,{\sc ii}]$\lambda$6584, 
[S\,{\sc ii}]$\lambda$6717+$\lambda$6731, 
[S\,{\sc iii}]$\lambda$9068. 
The intensities of all lines are normalised to the H$\beta$ line flux.  
The line intensity [O\,{\sc iii}]$\lambda$4959 is required to define the $R_3$ value. 
However this line is not reported in some of the papers considered here. Therefore the $R_3$ value 
is derived from the [O\,{\sc iii}]$\lambda$5007 line intensity (see below). 
Similarly, the values of $N_2$ and $S_3$ are estimated without the lines 
[N\,{\sc ii}]$\lambda$6548 and [S\,{\sc iii}]$\lambda$9532, which are also 
not reported in some papers.
Furthermore, only the summed-up line fluxes [S\,{\sc ii}]$\lambda$6717+$\lambda$6731 
are available in a number of publications. 

We have taken the de-reddened line intensities as reported by the authors.
In some papers the measured fluxes are reported only. 
In these cases, the measured emission-line fluxes were corrected for interstellar 
reddening using the theoretical H$\alpha$ to H$\beta$ ratio  (i.e., the standard  value 
of H$\alpha$/H$\beta$ = 2.86) and the 
analytical approximation to the Whitford interstellar reddening law 
from \citet{Izotov1994ApJ435}. 

As was noted above, only one line in a doublet
([O\,{\sc iii}]$\lambda$5007 from the doublet [O\,{\sc iii}]$\lambda$5007 + $\lambda$4959,  
[N\,{\sc ii}]$\lambda$6584 from the doublet [N\,{\sc ii}]$\lambda$6584 + $\lambda$6548,  
and [S\,{\sc iii}]$\lambda$9068 from the doublet [S\,{\sc iii}]$\lambda$9068 + $\lambda$9532) 
is given in some publications.
The [O\,{\sc iii}]$\lambda$5007 and $\lambda$4959 lines originate from transitions from the 
same energy level, so their flux ratio is due only to the transition probability ratio, 
which is  very close to 3 \citep{storey2000}. Therefore, the value of $R_3$ can be estimated as 
$R_3  = 1.33$[O\,{\sc iii}]$\lambda$5007.  
Similarly, the [N\,{\sc ii}]$\lambda$6584 and $\lambda$6548 lines also originate from transitions from the 
same energy level and the transition probability ratio for those lines is again
close to 3 \citep{storey2000}. The value of $N_2$ can therefore be estimated as 
$N_2$  = 1.33[N\,{\sc ii}]$\lambda$6584.
Furthermore, the transition probability ratio for 
[S\,{\sc iii}]$\lambda$9532 and  [S\,{\sc iii}]$\lambda$9068 is 2.44
\citep{mendoza1982}. Hence, the value of $S_3$ can be estimated as 
$S_3$  = 3.44[S\,{\sc iii}]$\lambda$9068.

The spectroscopic data so assembled form the basis of the present study. 
Our list contains  714 spectra. Since two or three auroral lines are detected 
in some spectra the resulting number of electron temperatures measurements is  899
(645 measurements of $t_{3,O}$ electron temperatures, 140 measurements of $t_{2,N}$ 
electron temperatures, and 114 measurements of $t_{3,S}$ electron temperatures).

\subsection{Abundance derivation}

In principle, the T$_{e}$ method, based on measurements of 
temperature-sensitive line ratios, should give accurate oxygen 
abundances. In practice, however, oxygen abundances in the same H\,{\sc ii} 
region derived by various authors can differ because
there may be errors in the line intensity measurements and the adopted atomic 
data may not be the same. 
To ensure that we have a relatively homogeneous data set, we have recalculated 
electron temperatures and oxygen and nitrogen abundances for all the 
H\,{\sc ii} regions. 

To convert the values of the line fluxes to the electron temperatures 
$t_{\rm 3,O}$, $t_{\rm 2,N}$, and $t_{\rm 3,S}$ and to the ion abundances O$^{++}$/H$^+$, 
O$^+$/H$^+$, and  N$^+$/H$^+$,  
we have solved the five-level atom model for the O$^{++}$, O$^+$, N$^+$, and S$^{++}$ ions,   
using recent atomic data.  
The Einstein coefficients for the spontaneous transitions $A_{jk}$   
for the five low-lying levels for all ions above have been taken from \cite{froese2004}.  
The energy level data are from \citet{edlen1985} for O$^{++}$,  
from \citet{wenaker1990} for O$^+$, from  \citet{galavis1997} for N$^+$, and from 
\citet{johansson1992}  for S$^{++}$.  
The effective cross sections (or effective collision strengths) for electron  
impact $\Omega$$_{jk}$ as a function of temperature are from \citet{aggarwal1999} for O$^{++}$,  
from \citet{pradhan2006} for O$^{+}$, from \citet{hudson2005} for N$^{+}$,   
and from \citet{tayal1999}  for S$^{++}$.
To derive the effective cross section for a given electron temperature, we have 
fitted a second-order polynomial to these data.

In the low density regime ($n_e \la 100$~cm$^{-3}$), the following simple  
expressions provide approximations to the numerical results with an accuracy 
better than 1\%. 
The electron temperatures are related to the measured line fluxes in the following way:    
\begin{equation} 
t_{\rm 3,O}   = \frac{1.467}{\log Q_{\rm 3,O} -0.876 - 
0.193\log t_{\rm 3,O} + 0.033\,t_{\rm 3,O} }
\label{equation:t3o}    
\end{equation} 
and
\begin{equation} 
t_{\rm 2,N}   = \frac{1.118}{\log Q_{\rm 2,N} -0.891 - 
0.177\log t_{\rm 2,N} + 0.030\,t_{\rm 2,N} } 
\label{equation:t2n}    
\end{equation} 
or
\begin{equation} 
t_{\rm 3,S}   = \frac{0.915}{\log Q_{\rm 3,S} -0.683 + 
0.485\log t_{\rm 3,S} - 0.114\,t_{\rm 3,S} } 
\label{equation:t2o}    
\end{equation} 
where $Q_{\rm 3,O}$ = 
[O\,{\sc iii}]($\lambda$4959+$\lambda$5007)/[O\,{\sc iii}]$\lambda 4363$ 
is the ratio of nebular to auroral oxygen O$^{++}$ line intensities, 
$Q_{\rm 2,N}$ =
[N\,{\sc ii}]($\lambda 6548+\lambda 6584$)/[N\,{\sc ii}]$\lambda 5755$ 
is the ratio of nebular to auroral nitrogen N$^{+}$ line intensities, and   
$Q_{\rm 3,S}$ =
[S\,{\sc iii}]($\lambda 9068+\lambda 9532$)/[S\,{\sc iii}]$\lambda$ 6312 
is the ratio of nebular to auroral sulphur S$^{++}$ line intensities.

The equations relating ion abundances to measured line fluxes are: 
\begin{equation} 
12+\log\frac{\rm N^+}{\rm H^+} = \log N_{2} + 6.263 + \frac{0.893}{t_2} - 
 0.603\,\log t_2 - 0.003\,t_2,
\label{equation:n2}    
\end{equation} 
\begin{equation} 
12+\log\frac{\rm O^+}{\rm H^+} = \log R_{2} + 5.929 + \frac{1.617}{t_2} - 
 0.568\,\log t_2 - 0.008\,t_2  ,
\label{equation:o2}    
\end{equation} 
and
\begin{equation} 
12+\log\frac{\rm O^{++}}{\rm H^+} = \log R_{3} + 6.251 + \frac{1.204}{t_3} - 
 0.613\,\log t_3 - 0.015\,t_3  .
\label{equation:o3}    
\end{equation} 
The total oxygen abundance is determined from 
\begin{equation}
\frac{\rm O}{\rm H} = \frac{\rm O^{++}}{\rm H^+} + \frac{\rm O^{+}}{\rm H^+}  . 
\label{equation:oh}   
\end{equation} 
In general, the small fraction of undetected O$^{3+}$ ions in the high-excitation 
 H\,{\sc ii} regions (O$^{+}$/(O$^{+}$+O$^{2+}$) $<$ 0.1) should be added to the oxygen
abundance \citep{Izotov2006AA448}. However, this results in only a minor correction to 
the oxygen abundance derived just from O$^{+}$ and O$^{2+}$. For example, the correction is around 
0.01 dex for the lowest-metallicity blue compact dwarf galaxy SBS 0335-052
\citep{Izotov2009AA503}. Hence, this correction is not considered in the following.

The total nitrogen abundance is determined from 
\begin{equation}
\log\frac{\rm N}{\rm H} = \log\frac{\rm O}{\rm H}  +  \log\frac{\rm N}{\rm O}
\label{equation:enh}   
\end{equation}
assuming  \citep{PeimbertCostero1969} 
\begin{equation}
\frac{\rm N^+}{\rm O^+} = \frac{\rm N}{\rm O}  .
\label{equation:xx}   
\end{equation}
The N$^+$/O$^+$ ion abundance ratio is derived from 
\begin{equation}
\log\frac{\rm N^+}{\rm O^+} = \log\frac{\rm N_2}{\rm R_2}  + 0.334 - 
\frac{0.724}{t_2} - 0.035 \log t_2 + 0.005 t_2 
\label{equation:eno}   
\end{equation}
which is obtained by combining Eq.(\ref{equation:n2}) with Eq.(\ref{equation:o2}). 

We have calculated electron temperatures and oxygen and nitrogen abundances for 
H\,{\sc ii} regions within the framework of the standard H\,{\sc ii} region model 
with two distinct temperature zones within the nebula. 
The electron temperature $t_3$ within the zone O$^{++}$ is given by the electron 
temperature $t_{3,O}$, and the temperature $t_2$ within the zones O$^{+}$ and N$^{+}$ 
is given by the temperature $t_{2,N}$.
It is common practice that the value of only the electron temperature is measured 
and the value of the other temperature is determined from the $t_2$ -- $t_3$ relation. 
The commonly used $t_2$ -- $t_3$ relation \citep{Campbell1986MNRAS223,garnett1992} 
\begin{equation}
t_2 = 0.7\,t_3 + 0.3 
\label{equation:t2t3}   
\end{equation}
is adopted here. When the value of the electron temperature $t_{3,S}$ is measured 
then the value of $t_3$ is obtained from the relation after \citet{garnett1992} 
\begin{equation}
t_3 = 0.83\,t_{3,S} + 0.17 .
\label{equation:t3t3s}   
\end{equation}
Two or three auroral lines are detected in some spectra and, consequently, two or
three electron temperatures ($t_{3,O}$,  $t_{2,N}$,  $t_{3,S}$) can be measured. 
For those  H\,{\sc ii} regions, two or three values of the electron temperature $t_3$ 
and oxygen and nitrogen abundances are determined.

\subsection{The $C$ method}

Suppose we have an observed H\,{\sc ii} region and a sample of reference H\,{\sc ii} regions.  
To find the counterpart for the H\,{\sc ii} region under study, we will compare not the 
four measured nebular lines $R_3$, $R_2$, $N_2$, and $S_2$ directly, but instead four 
other values that are expressed in terms of these line intensities: $P$ = $R_3$/($R_2$ + $R_3$) 
(excitation parameter), log$R_3$, log($N_2$/$R_2$), and log($S_2$/$R_2$).  
A linear combination of these values can serve as an indicator of the
metallicity in an H\,{\sc ii} region \citep{pilyugin2010ApJ720}.

We specify the difference between the spectrum of the studied H\,{\sc ii} region and the
spectrum of the $j$th H\,{\sc ii} region from the reference sample as 
\begin{eqnarray}
       \begin{array}{lll}
\Delta {\rm Sp}_{j} & =  &  [\frac {1}{4}((\Delta \log (R_3)_j)^2 + (\Delta P_j)^2       \\
          &  +   & (\Delta \log (N_2/R_2)_j)^2 + (\Delta  \log (S_2/R_2)_j)^2)]^{\frac {1}{2}} .  \\
     \end{array}
\label{equation:dsp}
\end{eqnarray}
The reference H\,{\sc ii} region with the smallest value of the $\Delta$Sp  will be 
considered as the counterpart for the investigated H\,{\sc ii} region. 
Then the oxygen and nitrogen abundances and electron temperature in the 
studied H\,{\sc ii} region can be assumed to be the same as those in 
its counterpart. 

However, oxygen and nitrogen abundances obtained in this manner can 
still have considerable errors for the following reasons. \\
1. The number of reference  H\,{\sc ii} regions is limited, especially at 
the high-metallicity end. Hence, in some cases there may be significant 
differences between abundances of the studied H\,{\sc ii} region and its 
counterpart. \\ 
2. In some cases the smallest value of $\Delta$Sp does not corresponds to 
the minimum difference in oxygen abundance. That would be the case if  
all the values of $P$, log$R_3$, log($N_2$/$R_2$), and log($S_2$/$R_2$) 
would change in proportion to the change of the abundance, and the calibration 
coefficients would be the same for all those values. This is indeed not the case. 
The coefficients are in fact functions of the metallicity
(for example, the coefficient for log$R_3$ even changes sign going from 
low-metallicity to high-metallicity H\,{\sc ii} regions). 
Furthermore, the difference in log($N_2$/$R_2$) values between two H\,{\sc ii} regions 
with the same oxygen abundance but different N/O abundance ratios can be 
larger than the difference between two H\,{\sc ii} regions 
with different oxygen abundances but similar nitrogen abundances. 
Thus, one may assume that the reference H\,{\sc ii} region with the smallest value 
of the $\Delta$Sp has a similar, but not necessarily the abundance closest to that of
the studied H\,{\sc ii} region. \\ 
3. Finally, the $T_e$-based  abundances for the reference H\,{\sc ii} region can 
of course involve some errors and uncertainties.

To overcome these problems and obtain more reliable estimates of 
abundances and electron temperatures, a number of reference H\,{\sc ii} regions 
with metallicities near the metallicity of the selected counterpart H\,{\sc ii} region,
(in the metallicity interval $\pm$(O/H)$_{int}$) can be used. 
Using a sufficient number of H\,{\sc ii} regions with metallicities in a suitable interval,   
one can obtain a linear expression 
for the oxygen abundance (or nitrogen abundance and electron temperature) 
of the form
\begin{equation}
Y  = a_0 + a_{1}\log R_3  + a_{2}P + a_{3}\log (N_2/R_2)  + a_{4}\log (S_2/R_2)  ,   
\label{equation:cn}
\end{equation}
where $Y = 12+\log$(O/H), or $Y = 12 + \log$(N/H), or $Y = t_3$. 
The oxygen and nitrogen abundances determined this way will in the following be referred 
to as (O/H)$_{C}$ and (N/H)$_{C}$, respectively.

\subsection{Selection of the reference H\,{\sc ii} regions}

The selection of reference H\,{\sc ii} regions is not a trivial task.
Here we use an approach that is based on the idea  
that if an H\,{\sc ii} region belongs to the fundamental sequence of 
the photoionised nebulae,
and its line fluxes are measured accurately, then the different 
methods, based on different emission lines, should yield similar  
physical characteristics (such as electron temperatures and 
abundances) of that object \citep{Thuan2010ApJ712}.

The $C$ method requires we first select a sample of reference H\,{\sc ii} regions from the collected data. 
The uncertainty of the oxygen abundance can be quantified by the discrepancy between the $C$-based and 
the $T_e$-based oxygen abundances $D_{O/H}$ = (O/H)$_{C}$  -- (O/H)$_{T_{e}}$. 
Similarly, the uncertainty in the nitrogen  abundance  can be quantified by $D_{N/H}$ = (N/H)$_{C}$  -- (N/H)$_{T_{e}}$.   
One may select reference H\,{\sc ii} regions where the discrepancies in the 
oxygen and nitrogen abundances are less than a fixed value of  $D_{O/H}^*$  and $D_{N/H}^*$. 
We use an iterative procedure to select a sample of reference  H\,{\sc ii} regions.
In the first step, we determine the oxygen (O/H)$_{C}$  and nitrogen (N/H)$_{C}$  abundances 
from  Eq.~(\ref{equation:cn}) for each H\,{\sc ii} region in our list
using all the other H\,{\sc ii} regions as the reference sample. Then we select a subsample of  
H\,{\sc ii} regions for which the absolute difference between the $C$-based 
and the $T_e$-based abundances (O/H)$_{C}$  -- (O/H)$_{T_{e}}$  
is less than  $D_{O/H}^*$ and the absolute difference 
 (N/H)$_{C}$ -- (N/H)$_{T_{e}}$ is less than  $D_{N/H}^*$.  
In the second step, we again determine 
 oxygen (O/H)$_{C}$ and nitrogen  (N/H)$_{C}$  abundances 
from  Eq.~(\ref{equation:cn})
for each H\,{\sc ii} region in our list
using a selected subsample of H\,{\sc ii} regions as the reference sample. 
We then select a new subsample of  
H\,{\sc ii} regions for which the difference between $C$-based and $T_e$-based 
abundances is less than $D_{O/H}^*$ and $D_{N/H}^*$, i.e., a new 
sample of reference H\,{\sc ii} regions is obtained.
The algorithm converges after a number (around ten) of iterations. 
As was noted above, two or three auroral lines are detected in some spectra and, consequently, two or
three electron temperatures ($t_{3,O}$,  $t_{2,N}$,  $t_{3,S}$) can be measured. 
For those H\,{\sc ii} regions more than one (i.e., two or three) oxygen and nitrogen abundances are determined. 
If these abundances, for a given spectrum, satisfy the 
selection criteria, we choose that one for which the difference between $C$-based and $T_e$-based 
oxygen abundances is the smallest.

In general, the selected sample of reference H\,{\sc ii} regions depends 
on three parameters: \\
1) the adopted interval of metallicity around the metallicity of the counterpart (O/H)$_{int}$ which defines 
   a subsample of reference H\,{\sc ii} regions used to derive the coefficients in  Eq.~(\ref{equation:cn}), \\
2) the adopted maximum value of $D_{O/H}^*$, i.e., of the discrepancy between $C$-based and $T_e$-based 
   oxygen abundances in the reference H\,{\sc ii} regions, \\
3) the adopted maximum value of $D_{N/H}^*$, i.e., of the discrepancy between $C$-based and $T_e$-based 
   nitrogen abundances in the reference H\,{\sc ii} regions. 

Examining the samples of reference H\,{\sc ii} regions selected with different combinations 
of the values of  $D_{O/H}^*$  (from an interval of 0.06 -- 0.12 dex),  $D_{N/H}^*$ (from an interval of 0.06 -- 0.12 dex) and  
(O/H)$_{int}$ (from an interval of 0.15 -- 0.30 dex), we find that oxygen and nitrogen abundances 
obtained by the $C$ method with $D_{O/H}^*$ in the range 0.08 to 0.12 dex,  $D_{N/H}^* = $ 0.08 to 0.12 dex, and  
(O/H)$_{int} = $ 0.20 to 0.30 dex are very similar. 

Therefore we will discuss below only three samples of the reference H\,{\sc ii} regions from 
the sequence defined by $D_{N/H}^* = D_{O/H}^*$ and (O/H)$_{int}  = 2D_{O/H}^*$. In this case each 
sample of reference H\,{\sc ii} regions can be specified by a single parameter,  $D_{O/H}^*$. 
The sample with  $D_{O/H}^* = 0.12$~dex has been selected from our compilation of 
 H\,{\sc ii} regions in the way described above. 
If the number of reference  H\,{\sc ii} regions within the adopted interval of metallicity around 
the metallicity of the counterpart (O/H)$_{int}$, which defines a subsample of reference H\,{\sc ii} regions 
used to derive the coefficients in Eq.(\ref{equation:cn}), is less than 12 
(this can occur at the high-metallicity end) then we increase the interval of metallicity (O/H)$_{int}$ 
with a step size of 0.05 dex until the number of reference  H\,{\sc ii} regions within the adopted interval of 
metallicity becomes larger than 12.   
This sample will be referred to as $Rsample12$ below. 
In a similar way the sample $Rsample10$ with  $D_{O/H}^*$ = 0.10 dex was selected from the 
sample $Rsample12$, and the sample $Rsample08$ with $D_{O/H}^*$ = 0.08 dex was selected 
from the sample $Rsample10$.

\begin{figure}
\resizebox{1.00\hsize}{!}{\includegraphics[angle=000]{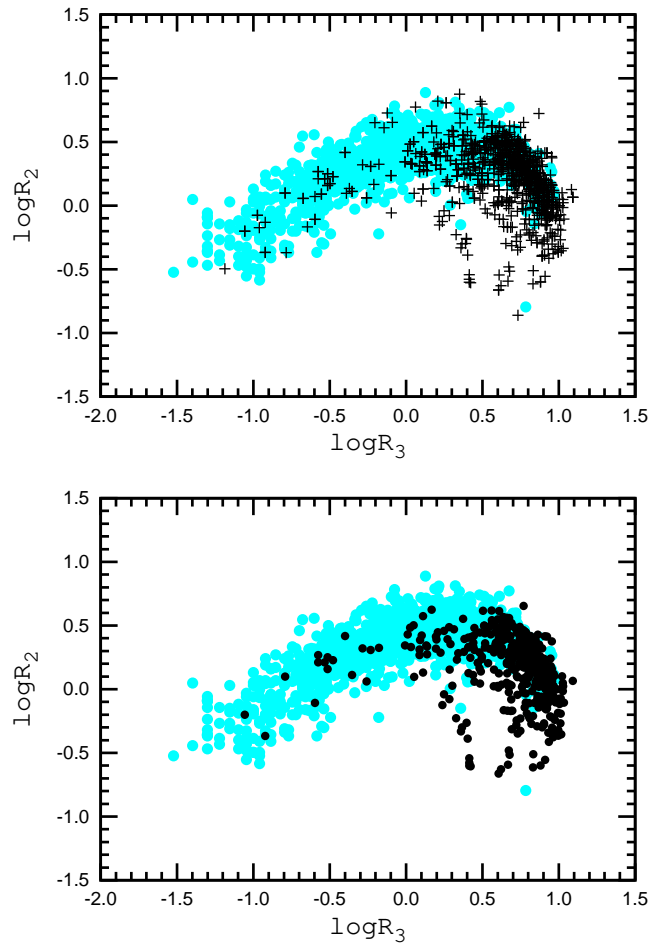}}
\caption{
The  log$R_{3}$ -- log$R_{2}$ diagram. 
The filled grey (light-blue in the color version) circles in both panels are 
H\,{\sc ii} regions from the compilation by \citet{Pilyugin2004AA425}. 
The dark (black) plus signs in the upper panel are 
H\,{\sc ii} regions from our present compilation.
The filled dark (black) circles in the lower panel show the selected reference  
H\,{\sc ii} regions, $Rsample10$.
(A color version of this figure is available in the online version.) 
}
\label{figure:lr3lr2}
\end{figure}

\begin{figure}
\resizebox{1.00\hsize}{!}{\includegraphics[angle=000]{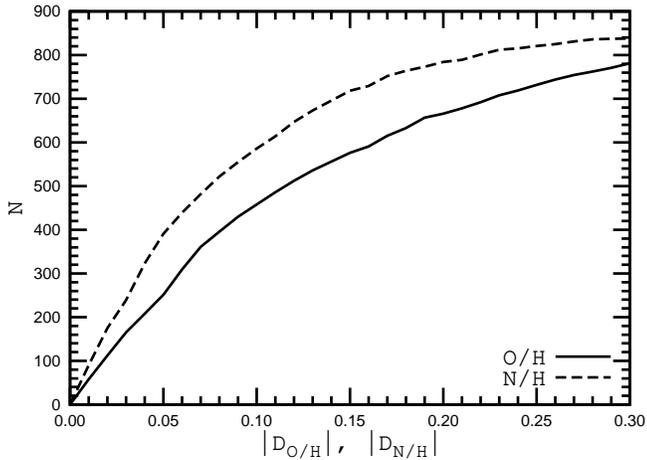}}
\caption{
The solid line shows the cumulative number of individual oxygen abundance 
measurements of H\,{\sc ii} regions with the absolute value of the 
difference between the $C$-based and the $T_e$-based abundances
(discrepancy index $D_{\rm O/H}$) less than a given value. 
The dashed line is the same but for nitrogen abundance.
}
\label{figure:gist}
\end{figure}

We choose $Rsample10$ for our abundance derivation. 
$Rsample12$ and $Rsample08$ will be used below to illustrate that the $C$-method 
abundances are robust. 
The reference H\,{\sc ii} regions from the $Rsample10$ are listed in 
Table~\ref{table:rsample}\footnote{The Table A1 is available in the electronic edition of the journal. 
The Table A1 is also publicly available in electronic form at 
{\tt http://dc.zah.uni-heidelberg.de/hiicounter/q/web/form}. 
The Fortran code for the  
determination of the oxygen and nitrogen abundances and electron temperature through 
the $C$ method (along with an example) is also available there.}. 
Column 1 in the Table~\ref{table:rsample} is the order number of the H\,{\sc ii} region. 
The de-reddened line intensities (in units of H$\beta$ line flux) are given in columns 2 to 5. 
$T_e$-based oxygen and nitrogen abundances [in units of 12+log(X/H)] are listed in columns 6 and 7 respectively, and
the electron temperature (in units of 10$^4$ K) is reported in column 8. 
In all cases where the electron temperature was reduced to $t_{3,O}$ [Eq.~(\ref{equation:t2t3}) or Eq.~(\ref{equation:t3t3s})] this is mentioned.
The index $j_T$ in column 9 is equal to 1 when 
the electron temperature $t_{3,O}$  is derived from the auroral line [O\,{\sc iii}]$\lambda$4363,
equal to 2 when the temperature $t_{2,N}$ is derived from the auroral line [N\,{\sc ii}]$\lambda$5755,
and equal to 3 when the temperature $t_{3,S}$ is derived from the auroral line [S\,{\sc iii}]$\lambda$6312. 
Commonly used catalogue names for each H\,{\sc ii} region and the sources of the spectral data are listed in columns 10 and 11, respectively.

\begin{figure}
\resizebox{1.00\hsize}{!}{\includegraphics[angle=000]{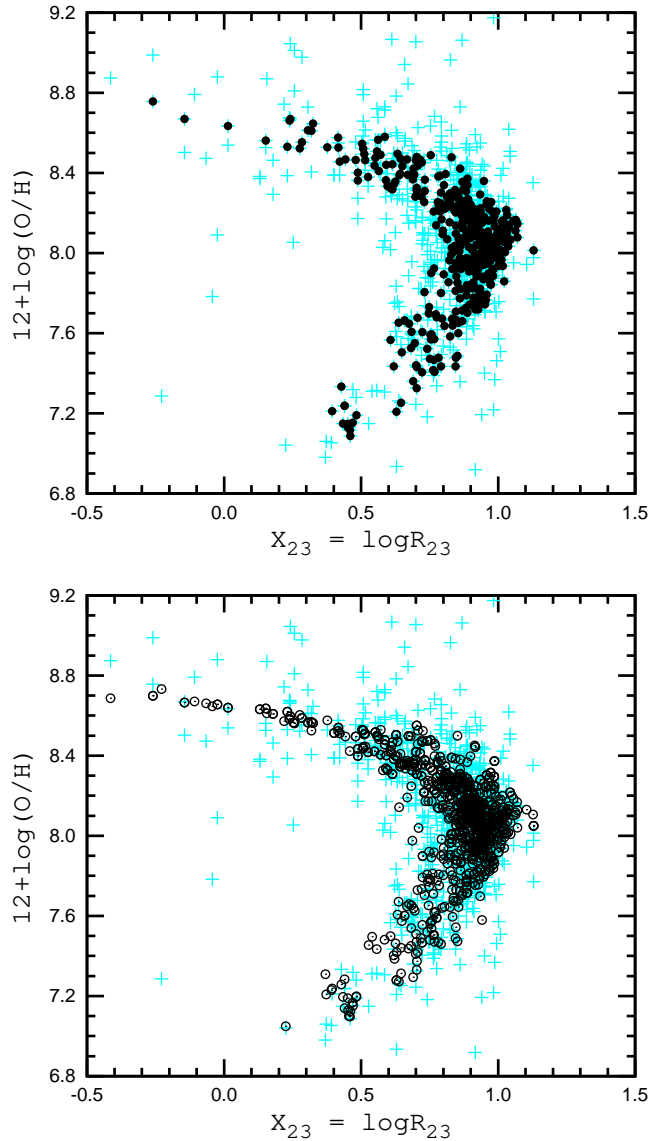}}
\caption{
The log$R_{23}$ -- O/H diagram. 
The grey (light-blue) plus signs in both panels are 
H\,{\sc ii} regions from our present compilation for the case of the $T_e$-based abundances. 
The dark (black) filled circles  in the upper panel show the selected reference  
H\,{\sc ii} regions, $Rsample10$.
The open dark (black) circles in the lower panel the H\,{\sc ii} regions 
from the present compilation for the case of the $C$-based oxygen abundances.
(A color version of this figure is available in the online version.) 
}
\label{figure:x23oh}
\end{figure}

\begin{figure}
\resizebox{1.00\hsize}{!}{\includegraphics[angle=000]{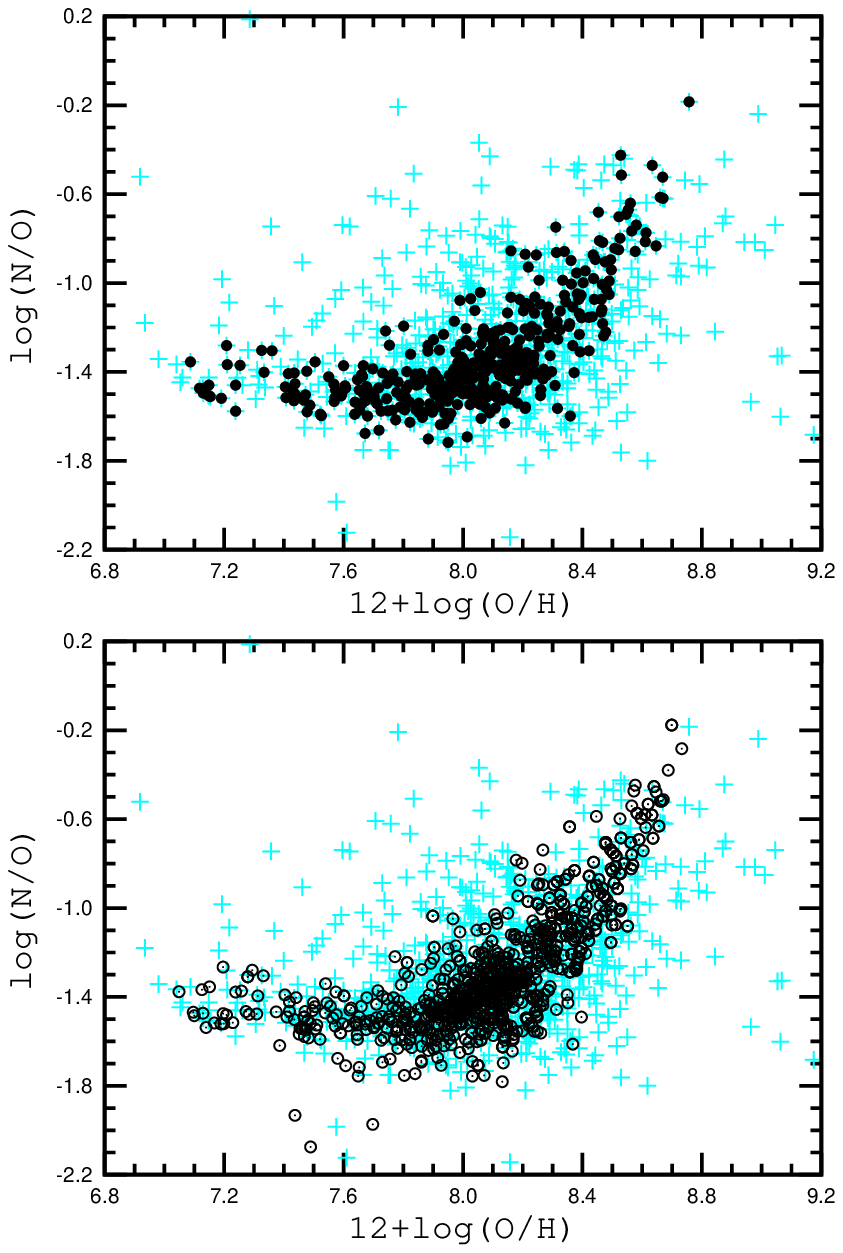}}
\caption{
The O/H -- N/O diagram. 
The grey (light-blue) plus signs in both panels are 
H\,{\sc ii} regions from our present compilation for the case of the $T_e$-based abundances. 
The dark (black) filled circles  in the upper panel show the selected reference  
H\,{\sc ii} regions, $Rsample10$.
The open dark (black) circles in the lower panel are the H\,{\sc ii} regions 
from the present compilation for the case of the $C$-based oxygen abundances.
(A color version of this figure is available in the online version.) 
}
\label{figure:ohno}
\end{figure}

Let us briefly consider the general properties of $Rsample10$. 
Fig.~\ref{figure:lr3lr2} shows the log$R_{3}$ vs.\ log$R_{2}$ diagram. 
\citet{Pilyugin2004AA425} compiled a 
large number of strong emission line measurements in spectra of individual 
H\,{\sc ii} regions in nearby spiral and irregular galaxies. 
Those H\,{\sc ii} regions are shown by filled gray (light-blue in the color version) circles in  
Fig.~\ref{figure:lr3lr2} (both panels) in order to outline the area occupied by  
 H\,{\sc ii} regions of nearby galaxies in the $R_{3}$ vs.\  $R_{2}$ diagram. 
The H\,{\sc ii} regions from the present compilation are shown by the dark (black) plus 
signs in the upper panel of Fig.~\ref{figure:lr3lr2}.     
The selected reference H\,{\sc ii} regions are shown by filled dark (black) circles  
in the lower panel of Fig.~\ref{figure:lr3lr2}. 

The solid line in Fig.~\ref{figure:gist} shows the cumulative number of individual oxygen abundance 
measurements from our compilation of H\,{\sc ii} regions with the absolute 
difference between the $C$-based and the $T_e$-based oxygen abundances
(discrepancy index D$_{\rm O/H}$) less than a given value. 
The dashed line shows the same for the nitrogen abundances.
One can clearly see that the cumulative numbers of nitrogen abundance  
measurements at any discrepancy index is larger than that for 
oxygen abundances, i.e., agreement between the $C$-based and the $T_e$-based abundances 
is better for nitrogen than for oxygen abundances.

Fig.~\ref{figure:x23oh} shows the log$R_{23}$ vs.\ O/H diagram. 
The $T_e$-based oxygen abundances for H\,{\sc ii} regions from our compilation are indicated by the grey (light-blue) 
plus signs in both panels of Fig.~\ref{figure:x23oh}.     
The log$R_{23}$ vs.\ O/H relation for the case of $C$-method-based oxygen abundances is shown by the open dark (black) circles 
in the lower panel of Fig.~\ref{figure:x23oh}. The adopted reference sample is shown by filled dark (black) circles  
in the upper panel of Fig.~\ref{figure:lr3lr2}. 

Fig.~\ref{figure:ohno} shows the O/H vs.\ N/O diagram. 
Again, the $T_e$-based abundances for H\,{\sc ii} regions in our compilation are shown by the grey (light-blue) 
plus signs in both panels of Fig.~\ref{figure:ohno}, and the reference H\,{\sc ii} regions are indicated by filled dark (black) circles  
in the upper panel of Fig.~\ref{figure:ohno}. 
The  O/H vs.\ N/O diagram for all the H\,{\sc ii} regions with $C$-based abundances is shown by the open dark (black) circles 
in the lower panel of Fig.~\ref{figure:ohno}. 

Analysing Fig.~\ref{figure:lr3lr2} -- Fig.~\ref{figure:ohno} we see that 
the number of the H\,{\sc ii} regions with metallicities 12+log(O/H) $\ga 8.4$ 
is small in the reference sample. More high-precision 
measurements of spectra of high-metallicity H\,{\sc ii} regions are obviously needed. 
 However, that requires measurements of extragalactic  H\,{\sc ii} regions using the largest telescopes. 
As was noted above, the Galactic  H\,{\sc ii} regions cannot be used as reference  H\,{\sc ii} 
regions because only parts of these H\,{\sc ii} regions are measured due to their large angular extent and 
therefore the obtained line intensities are usually not representative for 
the nebula as a whole.

\begin{figure*}
\resizebox{1.00\hsize}{!}{\includegraphics[angle=000]{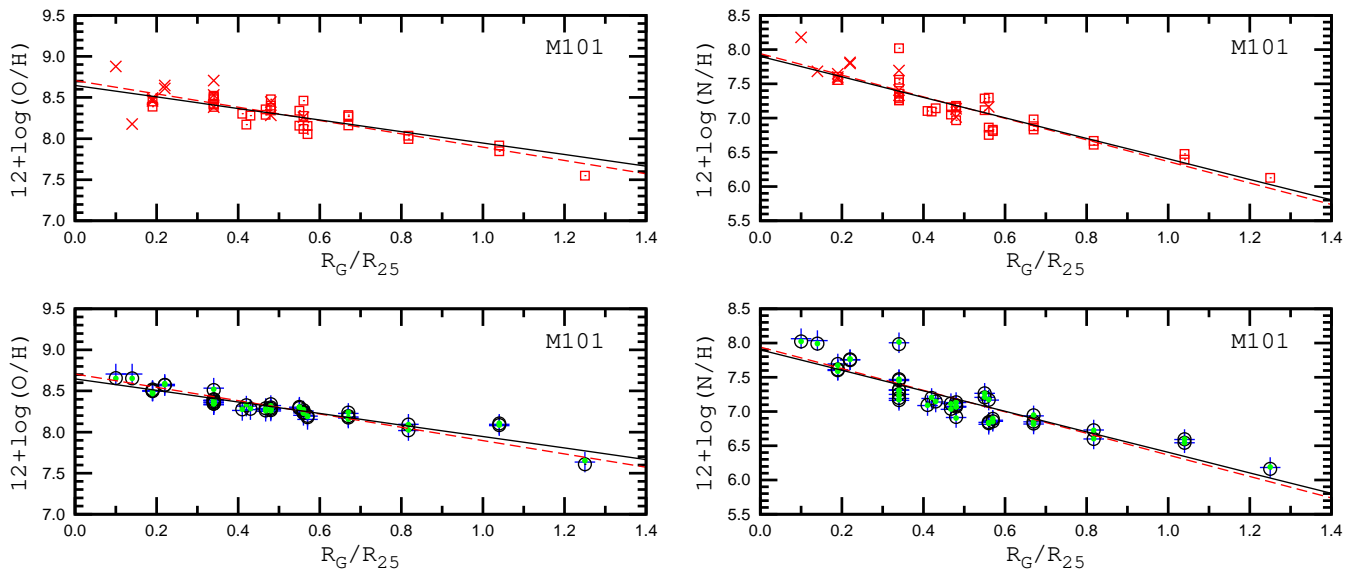}}
\caption{
The radial distribution of the oxygen (left column panels) and nitrogen (right column panels) 
in the disc of the spiral galaxy M~101.  In the upper two panels,
$T_e$-based abundances are shown by the (red) squares (with measured $t_{3,O}$) and 
the (red) crosses (with measured $t_{2,N}$). The linear best fit to these data points is shown
by the long-dashed (red) line. In the lower two panels, 
the $C$-based abundances obtained with $Rsample08$, $Rsample10$, and $Rsample12$ are shown by the (green) dots, 
open circles [the best fit to those data is given by the solid (black) line], and plus signs, respectively. 
}
\label{figure:m101}
\end{figure*}

\subsection{Verification of the $C$ method}

\begin{figure*}
\resizebox{1.00\hsize}{!}{\includegraphics[angle=000]{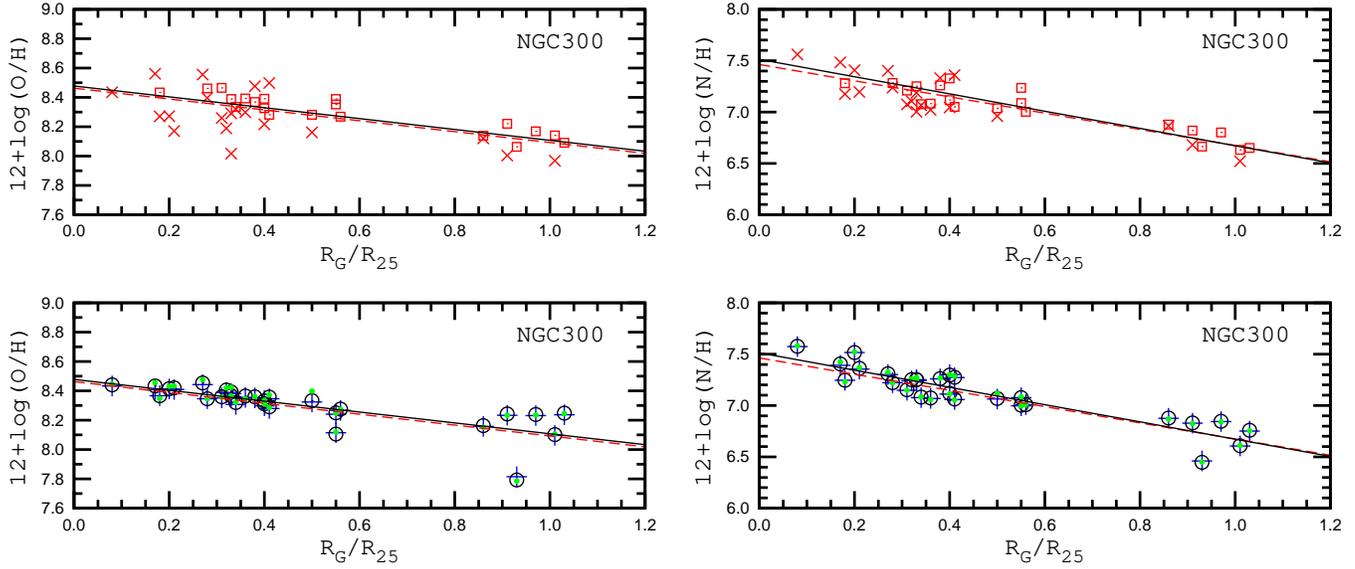}}
\caption{
Same as Fig.~\ref{figure:m101} but for the disc of the spiral 
galaxy NGC~300. 
}
\label{figure:ngc300}
\end{figure*}

\begin{figure*}
\resizebox{1.00\hsize}{!}{\includegraphics[angle=000]{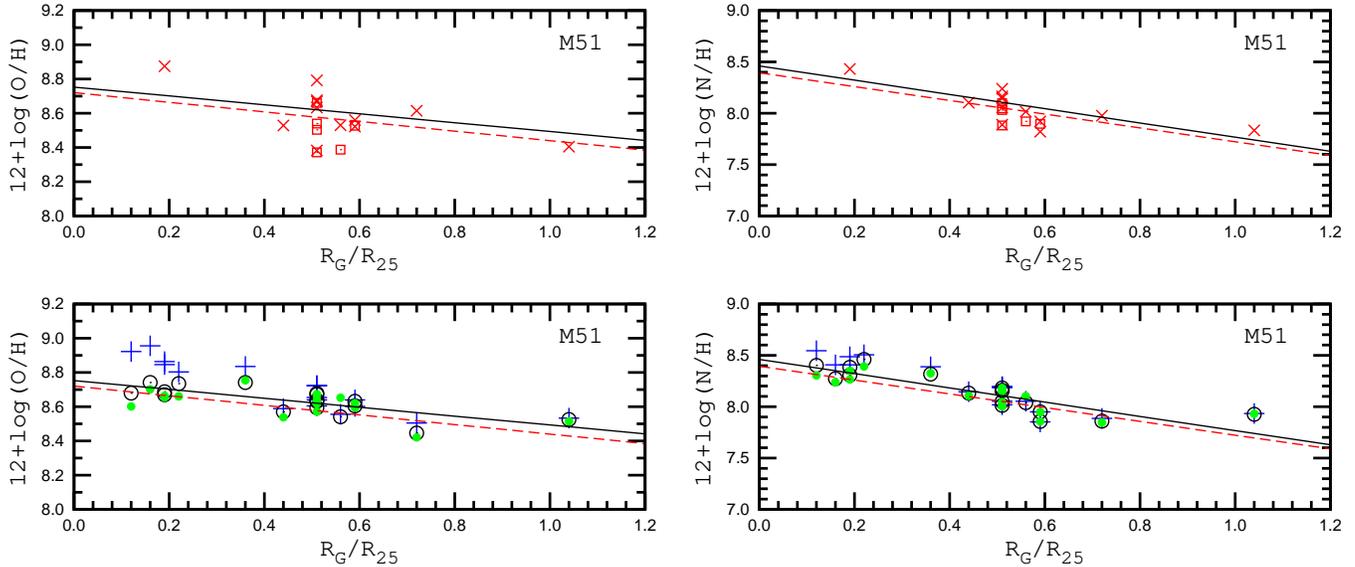}}
\caption{
Same as Fig. \ref{figure:m101} but for the disc of the spiral galaxy M~51.
The (red) squares are the $T_e$-based abundances with measured electron temperature $t_{3,S}$.
The other notations are the same as in  Fig.~\ref{figure:m101}.
}
\label{figure:m51}
\end{figure*}

The radial oxygen abundance distributions in the discs of the 
spiral galaxies M~101, NGC~300, and M~51 were derived based on H\,{\sc ii} regions with 
measured electron temperatures
\citep{Kennicutt2003ApJ591,Bresolin2009ApJ700,Bresolin2004ApJ615}, 
which offers a possibility to verify the $C$ method 
and different samples of reference H\,{\sc ii} regions by comparing the radial distribution of the 
$C$-based abundances with those traced by the H\,{\sc ii} regions with $T_e$-based 
abundances. The fractional radius, the galactocentric distance $R_G$ expressed in 
terms of the isophotal radius $R_{25}$, is used in the diagrams where radial  
oxygen (nitrogen) abundance distributions are plotted. 

In Fig.~\ref{figure:m101} we compare the radial distributions of the oxygen 
and nitrogen abundances obtained by the $C$ method using different samples of
reference H\,{\sc ii} regions with radial distributions traced by $T_e$-based 
abundances in the disc of the M~101 \citep[from the compilation by][]{Pilyugin2011MNRAS412}. 
The upper panels in Fig.~\ref{figure:m101} show the $T_e$-based 
oxygen (upper left panel) and nitrogen (upper right panel) abundances: 
the open (red) squares are H\,{\sc ii} regions with measured $t_{3,O}$ electron temperatures 
(detected [O\,{\sc iii}]$\lambda$4363 line) and the (red) crosses are those  
 with measured $t_{2,N}$ electron temperatures (detected 
[N\,{\sc ii}]$\lambda$5755). If both auroral lines are available then we derive
two values of oxygen (or nitrogen) abundances. The long-dashed (red) 
line is the best linear fit to these data. 
The symbols in the lower panels in Fig.~\ref{figure:m101} show the $C$-method based 
oxygen (lower left panel) and nitrogen (lower right panel) abundances from the same 
sample of H\,{\sc ii} regions. The (green) points, (blue) plus signs, and (black) open circles
show abundances obtained with the $Rsample08$, $Rsample10$, and $Rsample12$ reference 
samples, respectively.

Fig.~\ref{figure:m101} shows that the oxygen and nitrogen 
abundances determined by the $C$ method using the three different reference samples 
are in agreement with each other and the radial distributions 
of the $C$-method-based oxygen and nitrogen abundances follow to the same trends traced by 
 $T_e$-based abundances. Fig.~\ref{figure:ngc300} shows the same for the radial abundance distributions 
in the disc of the spiral galaxy NGC~300. The abundances of H\,{\sc ii} regions were determined 
 with the spectral measurements from \citet{Bresolin2009ApJ700}. Again, the oxygen and nitrogen 
abundances derived by the $C$ method using the three different reference samples are in agreement 
with each other, and the radial distributions 
of $C$-method-based oxygen and nitrogen abundances follow almost exactly the trends traced by 
 $T_e$-based abundances (the best linear fits to $T_e$-based and $C$-method-based abundances
coincide and cannot be distinguished in Fig.~\ref{figure:ngc300}). 
It should be noted that the scatter in the $C$-method-based abundances, at a given galactocentric 
distance, is even smaller than the one in the $T_e$-based abundances.

Fig.~\ref{figure:m51} compares the radial distributions 
of $T_e$-based and $C$-method-based oxygen and nitrogen abundances in 
the disc of the spiral galaxy M~51. 
The $t_{2,N}$ and $t_{3,S}$ electron temperatures have been measured in a number of 
 H\,{\sc ii} regions in the disc of M~51 \citep{Bresolin2004ApJ615,Garnett2004ApJ607}.
The oxygen abundance gradient derived by \citet{Bresolin2004ApJ615} is shown by the 
long-dashed (red) line in the upper left panel. 
$C$-method-based abundances for the \citet{Bresolin2004ApJ615} total sample are derived
including H\,{\sc ii} regions where auroral lines were not detected (their Table 6).
Fig.~\ref{figure:m51} confirms again that the oxygen and nitrogen 
abundances determined by the $C$ method using the three different reference samples 
are in agreement with each other and the radial distributions 
of $C$-method-based oxygen and nitrogen abundances follow to the trends traced by 
 $T_e$-based abundances. 
 
From the above, we draw two conclusions: \\
1. The oxygen and nitrogen abundances estimated by the $C$ method 
using the three different reference samples ($Rsample12$, $Rsample10$, and $Rsample08$) 
are in good agreement, i.e., the $C$-method-based abundances are robust. \\
2.  The $C$-method-based oxygen and nitrogen abundances are also in agreement with the 
 $T_e$-based abundances, i.e., the $C$ method produces reliable oxygen and nitrogen abundances.

\subsection{Uncertainty in the $C$-based abundances, caused by 
errors in line intensities}

\begin{figure*}
\resizebox{1.00\hsize}{!}{\includegraphics[angle=000]{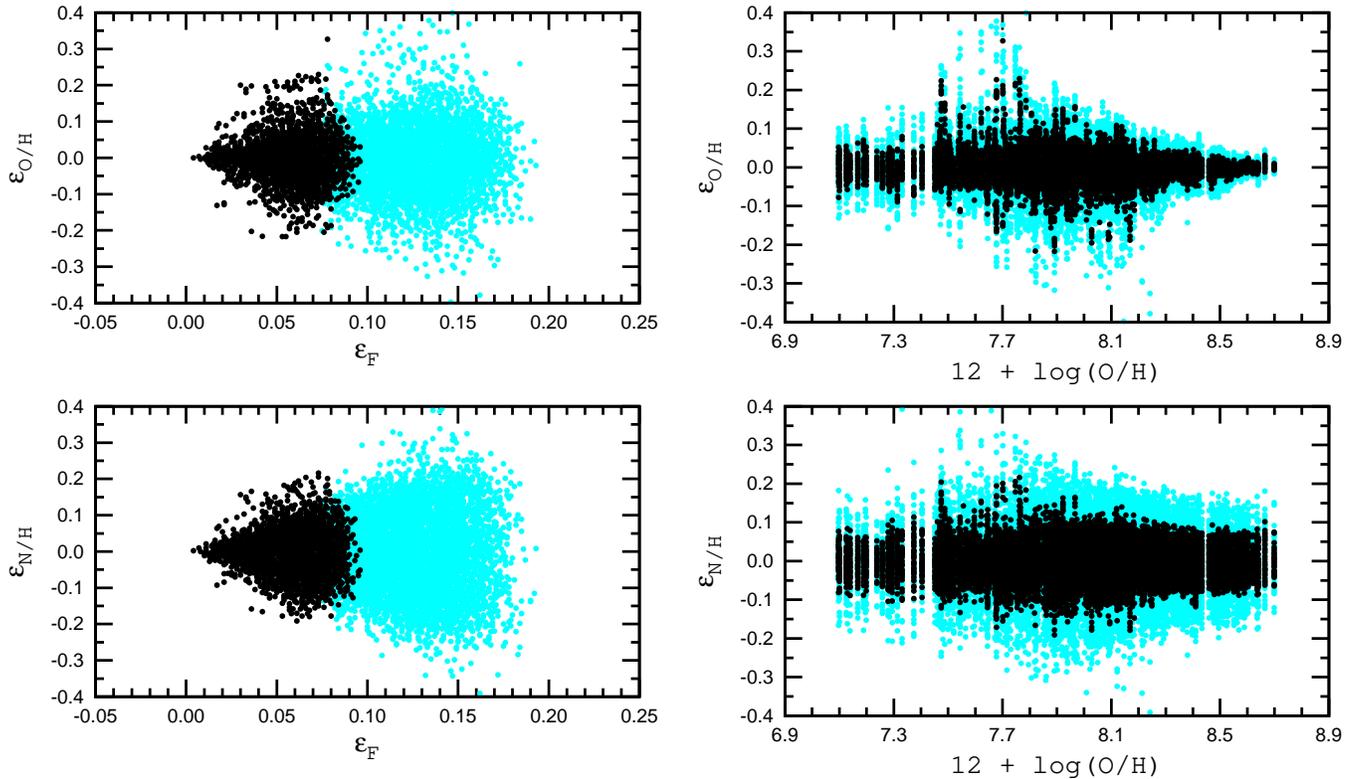}}
\caption{
The error in the oxygen $\epsilon$$_{\rm O/H}$ (upper-row panels) and 
nitrogen $\epsilon$$_{\rm N/H}$ (lower-row panels) abundances 
as a function of the error in the line fluxes $\epsilon _F$.(left-column panels) and 
oxygen abundance (right-column panels). 
The filled dark (black) circles show the Monte Carlo simulations with random 
relative error in each line intensity ranging from $-10$\% to $+10$\%. 
The filled grey (light-blue in the color version) circles show these 
 with random relative error ranging from $-20$\% to $+20$\%. 
(A color version of this figure is available in the online version.) 
}
\label{figure:ee}
\end{figure*}

The ``strong'' nebular lines $R_3$ and $R_2$ become rather weak in high-metallicity 
H\,{\sc ii} regions while the ``strong'' lines $N_2$ and $S_2$ are weak in low-metallicity 
H\,{\sc ii} regions. Therefore the measurements of those lines can have significant errors. 
The uncertainty in oxygen and nitrogen abundances, caused by the  
uncertainty in the line flux measurements, can be estimated in the following way.  
We consider the measured fluxes in the reference H\,{\sc ii} regions as the ``true'' 
fluxes and $C$-method-based oxygen and nitrogen abundances as the ``true'' abundances. 
We then introduce a random relative error $\epsilon$ to each and every line flux,
\begin{eqnarray}
       \begin{array}{lll}
R_{3}^{*} & = & (1 + \epsilon _{R_{3}}) \;  R_{3}^{\rm true},   \\
         &   &                                            \\
R_{2}^{*} & = & (1 + \epsilon _{R_{2}}) \;  R_{2}^{\rm true},  \\ 
         &   &                                            \\
N_{2}^{*} & = & (1 + \epsilon _{N_{2}}) \;  N_{2}^{\rm true},  \\ 
         &   &                                            \\
S_{2}^{*} & = & (1 + \epsilon _{S_{2}}) \;  S_{2}^{\rm true},   \\ 
     \end{array}
\label{equation:fluxe}
\end{eqnarray}
where $\epsilon_{R_3}$, $\epsilon_{R_2}$, $\epsilon_{N_2}$, and $\epsilon_{S_2}$ 
are produced by a random number generator.
We then determine the oxygen (O/H)$_C^*$  and nitrogen (N/H)$_C^*$ abundances based on the 
$R_{3}^{*}$, $R_{2}^{*}$, $N_{2}^{*}$, and $S_{2}^{*}$ line fluxes using the $C$ method. 
The differences log(O/H)$_C^* - \log$(O/H)$^{\rm true}$ and log(N/H)$_C^* - \log$(N/H)$^{\rm true}$ 
can be seen as a measure of the uncertainty in the oxygen and nitrogen abundances caused by the  
uncertainty in the line flux measurements. 

We have considered two cases. In the first, we adopt random relative errors 
$\epsilon _{R_{3}}$, $\epsilon _{R_{2}}$, $\epsilon _{N_{2}}$, and $\epsilon _{S_{2}}$ 
in the range of $-10$\% to $+10$\%. The mean error in the flux measurements is defined as
\begin{equation} 
\epsilon _F   = \sqrt{\frac{1}{4}(\epsilon_{R_3}^2 + \epsilon_{R_2}^2 + \epsilon_{N_2}^2 + \epsilon_{S_2}^2)}.    
\label{equation:ef}    
\end{equation} 
The (O/H)$_C^*$ and (N/H)$_C^*$ abundances for each reference H\,{\sc ii} region were computed 
with 50 different random errors added to each line flux. 
The values of the error in the oxygen and nitrogen abundances are defined as  
$\epsilon$$_{\rm O/H}$ = log(O/H)$_C^*$ -- log(O/H)$^{\rm true}$ and 
$\epsilon$$_{\rm N/H}$ = log(N/H)$_C^*$ -- log(N/H)$^{\rm true}$, respectively.
In the second case, we adopt random relative errors 
$\epsilon _{R_{3}}$, $\epsilon _{R_{2}}$, $\epsilon _{N_{2}}$ and $\epsilon _{S_{2}}$ 
in the range $-20$\% to $+20$\%.
 
The left panels in Fig.~\ref{figure:ee} show the errors in the oxygen $\epsilon$$_{\rm O/H}$
(upper panel) and nitrogen $\epsilon$$_{\rm N/H}$ (lower panel) abundances as a function of the mean error in 
the line fluxes $\epsilon _F$. The Monte Carlo simulations for the first case are shown by 
the dark (black) points, and those for the 
second case are represented by the grey (light-blue) points. 
In the right panels in Fig.~\ref{figure:ee} we plot the error in the oxygen $\epsilon$$_{\rm O/H}$
(upper panel) and nitrogen $\epsilon$$_{\rm N/H}$ (lower panel) abundances against the oxygen abundance. 

Fig.~\ref{figure:ee} reveals that the uncertainties in the oxygen 
abundances caused by the errors in the first case are not in excess of 0.1 dex for the vast majority of
H\,{\sc ii} regions, although in a few cases they may be larger. 
The uncertainties in the nitrogen abundances are slightly larger, up to 0.15 dex. 
It is interesting to note that even when the mean error in 
the line fluxes $\epsilon _F$ is relatively large 
the error in the oxygen $\epsilon$$_{\rm O/H}$
and nitrogen $\epsilon$$_{\rm N/H}$ abundances can be close to zero. 
This is because the errors in the abundances caused by errors in different line intensities 
can have opposite signs and therefore cancel, 
i.e., even when the error in each term on the right-hand side of Eq.~(\ref{equation:cn}) is considerable, 
the sum (abundance) can be correct if the errors in different right-hand side terms have opposite signs and 
compensate each other.

A closer look at the upper right panel in Fig.~\ref{figure:ee} shows that the uncertainty in the oxygen 
abundance caused by errors in the line intensities reach a maximum in H\,{\sc ii} regions with 
metallicities in the range from $12 + \log$(O/H) $\sim 7.8$ to $12 + \log$(O/H) $\sim 8.2$. 
This is because the strong line fluxes are less sensitive to the oxygen 
abundance of the H\,{\sc ii} regions this metallicity interval   
(in particular, the transition from the upper to the lower branch of the R$_{23}$ vs.\ O/H 
diagram occurs in this interval). 
At high metallicities 
($12 + \log$(O/H) $\ga 8.2$) the uncertainty decreases with increasing metallicity. 
This is due to the fact that the strong line fluxes change much more with the oxygen 
abundance in high-metallicity (cold)  H\,{\sc ii} regions than in low-metallicity (warm and hot) H\,{\sc ii} regions
\citep{pilyugin2010ApJ720}. 
In other words, similar relative changes in the strong line fluxes correspond to a smaller change 
in the oxygen abundance in high-metallicity H\,{\sc ii} regions than in low-metallicity H\,{\sc ii} regions. 
Hence, similar relative errors in the line fluxes result in smaller errors at high metallicity than at low metallicity.

\section{Application of the $C$ method: Abundance gradients in the extended discs of spiral galaxies}

Spectra of  H\,{\sc ii} regions in the outer disc of the spiral
galaxy M~83 (=NGC~5236) \citep{Bresolin2009ApJ695} and in the extended disc of 
the spiral galaxy NGC~4625 \citep{Goddard2011MNRAS412} were obtained quite recently.  
The abundance gradients were determined out to around 2.5 times the 
optical isophotal radius. It was found that at the transition between the inner and outer 
disc the abundance gradient becomes flatter. In addition, there appears to be an 
abundance discontinuity close to this transition. 
However, the abundances estimated with different calibrations differ by more than a factor of three. 
This prevents one from drawing a solid conclusion on the real behavior of the abundance gradients 
in the outer discs of these galaxies.

Oxygen abundance gradients have been obtained for a large sample of spiral galaxies  
\citep[][among others]{Vila1992MNRAS259,Zaritsky1992ApJ390,vanZee1998AJ116}. 
It was found that nearly all the gradients are reasonably well fitted by a
single exponential profile, although in several cases the gradient slope may
not be constant across the disc but instead flattens (or steepens) in the
outer disc.
In particular, a break in the abundance gradient of 
M~101 at $R_G$/$R_{25}$ $\sim$ 0.5 was reported quite early on \citep{Vila1992MNRAS259,Zaritsky1992ApJ390}. 
Oxygen abundances obtained with the strong-line method were used in those works. 
However, the $T_e$-based oxygen abundance distribution does not show the flattening
in the outer disc of M~101, (upper left panel in Fig.\ref{figure:m101}).
It was noted in the introduction that the strong-line relations 
may not work across the whole range of observed metallicities in H\,{\sc ii} regions. 
An unjustified use of the relationship between oxygen abundance and 
strong line intensities, constructed for high-metallicity H\,{\sc ii} regions of 
the upper branch of the R$_{23}$ -- O/H diagram, when determining
oxygen abundances in low-metallicity H\,{\sc ii} regions at the periphery of a galaxy, 
results in erroneous (overestimated) oxygen abundances, and, 
as a consequence, an erroneous bend in the slope of the abundance gradient 
\citep{Pilyugin2003AA397a}. 

\begin{figure}
\resizebox{1.00\hsize}{!}{\includegraphics[angle=000]{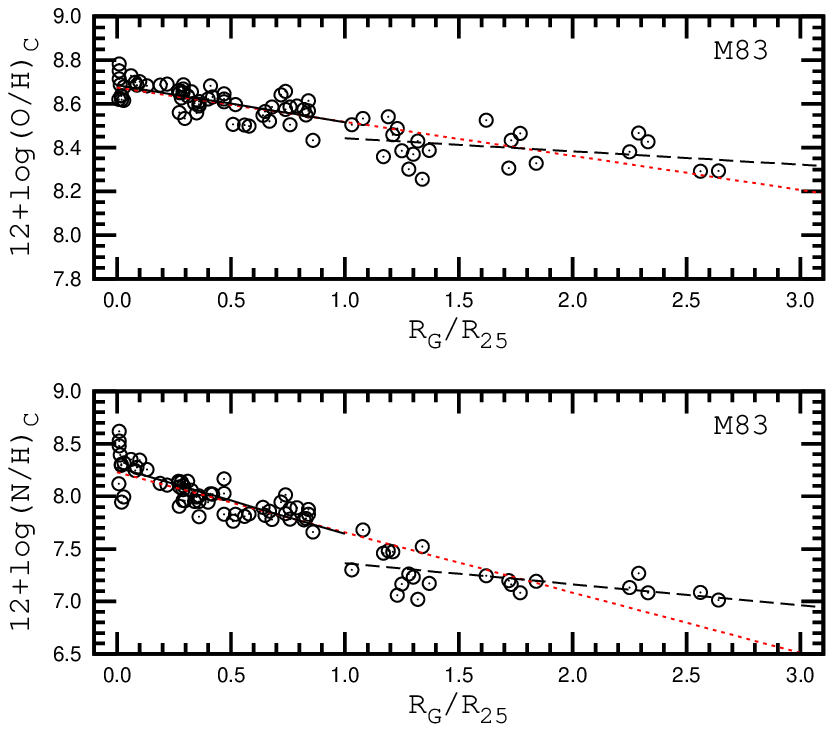}}
\caption{
The radial distributions of oxygen (upper panel) and nitrogen (lower panel) abundances 
in the extended disc of the spiral galaxy M~83. The abundances are estimated through 
the $C$ method.
The open circles stand for H\,{\sc ii} regions from 
\citet{Dufour1980ApJ236,Webster1983MNRAS204,Bresolin2002ApJ572,Bresolin2005AA441,Bresolin2009ApJ695,Esteban2009ApJ700}. 
The solid line is the best linear fit to the data points with galactocentric distances 
smaller the isophotal $R_{25}$ radius, the dashed line is the best linear fit to data points marking the H\,{\sc ii} regions
beyond the isophotal radius.
The dotted (red) line shows the best single linear fit to all the data points. 
(A color version of this figure is available in the online version.) 
}
\label{figure:m83}
\end{figure}

\begin{figure}
\resizebox{1.00\hsize}{!}{\includegraphics[angle=000]{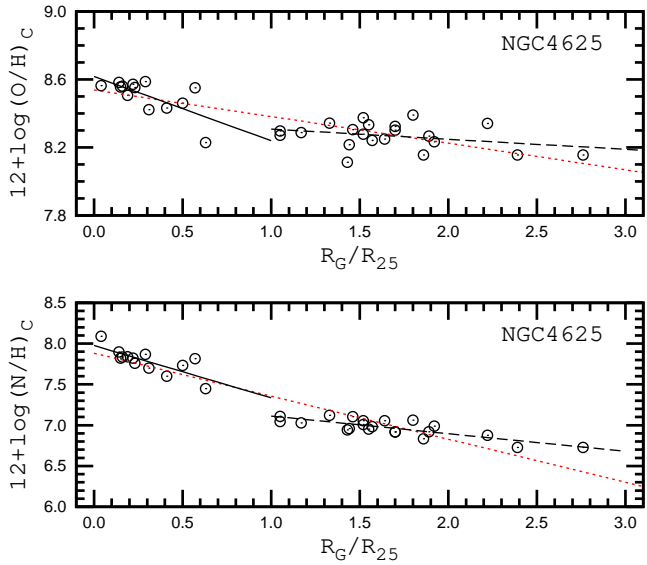}}
\caption{
Same as Fig. \ref{figure:m83} but for the spiral galaxy NGC~4625. 
The circles stand for the abundances in H\,{\sc ii} regions from \citet{Goddard2011MNRAS412} 
estimated through the $C$ method.
(A color version of this figure is available in the online version.) 
}
\label{figure:ngc4625}
\end{figure}

\begin{figure}
\resizebox{1.00\hsize}{!}{\includegraphics[angle=000]{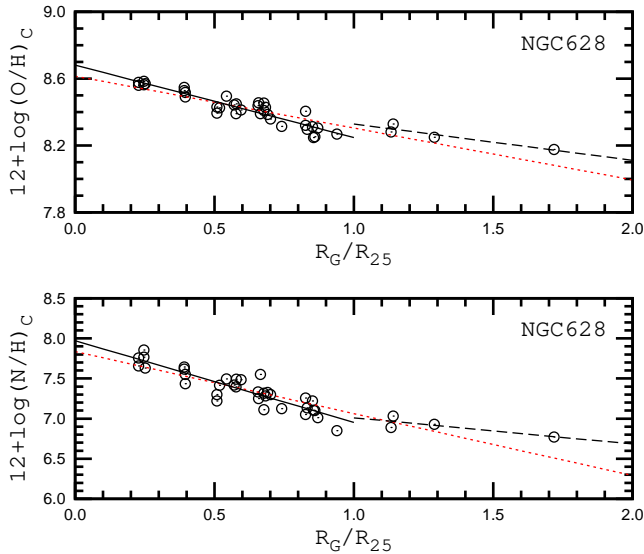}}
\caption{
Same as Fig.~\ref{figure:m83} but for the spiral galaxy NGC~628. 
The circles represent abundances in H\,{\sc ii} regions from 
\citet{McCall1985ApJS57,Ferguson1998AJ116,vanZee1998AJ116,Bresolin1999ApJ540} 
estimated through the $C$ method.
(A color version of this figure is available in the online version.) 
}
\label{figure:ngc0628}
\end{figure}

The $C$-method-based oxygen abundance distributions follow well the gradients traced 
by the $T_e$-based oxygen abundances, see Figs.~\ref{figure:m101} -- \ref{figure:m51}.  
Here we apply the $C$ method to derive abundance gradients in the extended 
discs of the galaxies M~83, NGC~4652 and NGC~628. In previous works, 
no attention was paid to the radial distribution of nitrogen 
abundances in the extended discs of those galaxies, despite the fact that such studies would have 
several advantages \citep{Thuan2010ApJ712}.  
First, since at $12+\log$(O/H) $\ga 8.3$, 
secondary nitrogen becomes dominant and the nitrogen abundance increases at 
a faster rate than the oxygen abundance \citep{Henry2000ApJ541}, 
the change in nitrogen 
abundances with galactocentric distance should show a larger amplitude in comparison to 
oxygen abundances and, as a consequence, the change of the gradient and/or 
abundance discontinuity should be easier to detect. 
Furthermore, there is a time delay in the nitrogen production as compared 
to oxygen production \citep{Maeder1992AA264,vandenhoek1997AAS123,Pagel1997book,Pilyugin2011ApJ726}. 
This provides an additional constraint on the chemical evolution of galaxies. 
These reasons led us to consider here not only  
the radial distribution of oxygen abundances but also that of 
nitrogen abundances. 

Fig.~\ref{figure:m83} shows the radial distributions of the $C$-method-based oxygen and nitrogen abundances in the disc 
of the spiral galaxy M~83, where line measurements were taken from 
\citet{Dufour1980ApJ236,Webster1983MNRAS204,Bresolin2002ApJ572,Bresolin2005AA441,Bresolin2009ApJ695,Esteban2009ApJ700}. 
The solid line is the best linear fit to the data points with galactocentric distances 
smaller than the isophotal $R_{25}$ radius, and the dashed line is for the H\,{\sc ii} regions
beyond the isophotal radius.
The dotted (red) line shows the best single linear fit to all the data points. 
Fig.~\ref{figure:ngc4625} shows the radial distributions of oxygen and nitrogen abundances in the disc 
of the spiral galaxy NGC~4625 for a sample of H\,{\sc ii} regions from \citet{Goddard2011MNRAS412}. 
Fig.~\ref{figure:ngc0628} shows the radial distributions of the oxygen and nitrogen abundances in the disc 
of the spiral galaxy NGC~628 for H\,{\sc ii} regions from 
\citet{McCall1985ApJS57,Ferguson1998AJ116,vanZee1998AJ116,Bresolin1999ApJ540}. 

Figs.~\ref{figure:m83},~\ref{figure:ngc4625}, and \ref{figure:ngc0628} demonstrate
that the gradient slopes within and beyond the optical isophotal radius are different. The gradient 
in the outer extended disc is shallower than that in the inner part of the galaxies.  
Thus, we confirm the conclusion of  \citet{Bresolin2009ApJ695} and  \citet{Goddard2011MNRAS412} that 
at the transition between the inner and outer disc the abundance gradient becomes flatter. 
It should be noted that the change in the gradient slope is more distinct in the radial distribution 
of nitrogen than oxygen abundances.
This is not surprising since the change in nitrogen 
abundances with galactocentric distance shows a larger amplitude in comparison to 
oxygen abundances and, as a consequence, the change of the gradient slope is easier to detect. 

On the other hand, Figs.~\ref{figure:m83},~\ref{figure:ngc4625}, and~\ref{figure:ngc0628} 
do not provide a solid case in favour of the existence of an abundance discontinuity in the transition from 
the inner to outer disc, as was suggested by \citet{Bresolin2009ApJ695} and \citet{Goddard2011MNRAS412}.  
Even if this abundance discontinuity exists, its amplitude is not in excess 
of the scatter in abundances among H\,{\sc ii} regions with similar galactocentric distances.

\section{Summary and conclusions}

In this paper, a new way of determining abundances and electron temperatures in H\,{\sc ii} 
regions based on strong emission lines is suggested. Our approach is based on the standard assumption
that H\,{\sc ii} regions with similar strong-line intensities have similar physical properties and abundances. 
A sample of reference H\,{\sc ii} regions with well-measured abundances is obtained, from which 
we choose a counterpart for the considered H\,{\sc ii} region by comparison of combinations of strong-line intensities.   
The oxygen and nitrogen abundances, as well as the electron temperature in the 
studied H\,{\sc ii} region may then be assumed to be the same as that in its counterpart.  
In other words, we suggest a method where abundances in H\,{\sc ii} regions are obtained ``by precedent''. 

To get more reliable abundances, we select a number of reference H\,{\sc ii} regions with abundances near those
in the counterpart H\,{\sc ii} region and then derive the abundance in the studied H\,{\sc ii} region through extra-/interpolation. 
We call this the counterpart method or, for brevity, the $C$ method. 

We have carried out an extensive search of the literature to compile a list 
of individual spectra of H\,{\sc ii} regions in irregular and 
spiral galaxies, with the requirement that they include the 
[O\,{\sc ii}]$\lambda$3727+$\lambda$3729, 
[O\,{\sc iii}]$\lambda$5007,
[N\,{\sc ii}]$\lambda$6584, 
[S\,{\sc ii}]$\lambda$6717+$\lambda$6731 lines  
and a detected auroral line of, at least, one ion. 
The spectroscopic data so assembled form the basis of the present study.
Our list contains 714 spectra. Since two or three auroral lines are detected 
in some spectra the total number of the electron temperature measurements is 
equal to 899.
To ensure that we have a relatively homogeneous data set, we recalculated 
electron temperatures and oxygen and nitrogen abundances for all the 
H\,{\sc ii} regions. 
Then we selected a sample of the reference H\,{\sc ii} regions from the collected data.
The list of our reference H\,{\sc ii} regions  ($Rsample10$) contains 414 objects.  

To verify the $C$ method we considered the radial distributions of the 
oxygen and nitrogen abundances in the discs of the spiral galaxies M~101, NGC~300, and M~51 for which abundance 
gradients were established on the basis of observed H\,{\sc ii} regions with measured electron temperatures. 
We found that the radial abundance gradients in the discs of these galaxies, 
as obtained by the $C$ method and the $T_e$ method, are in satisfactory agreement. 
This is evidence in favour of the $C$ method producing reliable abundances. 
Thus, the strong lines  
[O\,{\sc ii}]$\lambda$3727+$\lambda$3729, 
[O\,{\sc iii}]$\lambda$5007,
[N\,{\sc ii}]$\lambda$6584, 
and [S\,{\sc ii}]$\lambda$6717+$\lambda$6731 
allow us to estimate the oxygen and nitrogen abundances in H\,{\sc ii} regions using
the $C$ method and the resultant abundances are compatible with the $T_e$-based abundances.
If the errors in the line measurements are within 10\% then one can expect that the 
uncertainty in the $C$-based abundances are not in excess of 0.1 dex.   

Finally, the $C$ method has been applied to study the radial abundance distributions in 
the extended discs of the spiral galaxies M~83, NGC~4625, and NGC~628, which have been suggested to
show shallower oxygen abundance gradients in the outer disc (beyond the isophotal radius) 
than in the inner disc, and to investigate a discontinuity in the gradient that occurs in 
proximity of the optical edge of the galaxy. 
We confirm the conclusion of \citet{Bresolin2009ApJ695} and \citet{Goddard2011MNRAS412} that
the abundance gradient becomes flatter at the transition between the inner and outer disc. 
We found that the change in the gradient slope is more distinct in the radial distribution 
of nitrogen than of oxygen abundances, which is expected.
On the other hand, we do not find solid evidence for the existence of an abundance discontinuity at the transition from 
the inner to the outer disc as found by \citet{Bresolin2009ApJ695} and \citet{Goddard2011MNRAS412}.  
Even if this abundance discontinuity is real its amplitude is not in excess 
of the scatter in abundances among H\,{\sc ii} regions with similar galactocentric distances.

\section*{Acknowledgements}

We are grateful to the referee, \'{A}.R.\ L\'{o}pez-S\'{a}nchez, for his constructive comments. 
L.S.P.\ and E.K.G.\ acknowledge support within the framework of Sonderforschungsbereich 
(SFB 881) on ``The Milky Way System'' (especially subproject A5), which is funded by the German 
Research Foundation (DFG). 
L.S.P. thanks the hospitality of the Astronomisches Rechen-Institut at the 
Universit\"{a}t Heidelberg  where this investigation was carried out.
The Dark Cosmology Centre is funded by the Danish National Research Foundation.

\appendix

\section{Online material. Table A1.}

Table~\ref{table:rsample} contains the dereddened line intensities, 
oxygen and nitrogen abundances and 
the electron temperatures of the reference H\,{\sc ii} regions.

\setcounter{table}{0}
\begin{table*}
\caption[]{\label{table:rsample}
List of the reference H\,{\sc ii} regions. The order number of each object is shown in 
column 1. 
The dereddened line intensities 
(in units of H$\beta$ line flux) are given in columns 2 to 5. The $T_e$--based oxygen and 
nitrogen abundances (in units of $12+\log$(X/H)) are listed in columns 6 and 7. The electron 
temperature (in units of 10$^4$ K) is reported in column 8. The index $j_T$ in column 9 is equal to 1 when 
the electron temperature is derived from the auroral and nebular lines of [O\,{\sc iii}],
equal to 2 when the temperature is derived from the lines of [N\,{\sc ii}], and equal 
to 3 when the temperature is derived from the lines of [S\,{\sc iii}]. (The electron 
temperature reduced to $t_3$([O\,{\sc iii}]) is reported in all cases.) 
The  H\,{\sc ii} region identifier and the literature sources of the spectral data are 
listed in columns 10 and 11.
}
\begin{center}
\begin{tabular}{rrrrrccccll} \hline \hline
n                     &
logR$_3$              &
logR$_2$              &
logN$_2$              &
logS$_2$              &
12+log(O/H)           &
12+log(N/H)           &
t$_3$                 &
j$_T$                 &
H\,{\sc ii} region            &
reference            \\ \hline
  1 &  -0.793 &   0.100 &   0.173 &  -0.337 &   8.56 &   7.92 &   0.55 &  2 &    M51 CCM 10            &   \citet{Bresolin2004ApJ615}     \\ 
  2 &  -0.354 &   0.111 &   0.203 &  -0.387 &   8.66 &   8.05 &   0.56 &  3 &    M51 CCM 53            &   \citet{Bresolin2004ApJ615}     \\ 
  3 &  -0.258 &   0.061 &   0.210 &  -0.201 &   8.53 &   8.02 &   0.60 &  2 &    M51 CCM 54            &   \citet{Bresolin2004ApJ615}     \\ 
  4 &  -0.597 &  -0.108 &   0.192 &  -0.387 &   8.63 &   8.16 &   0.50 &  2 &    M51 CCM 55            &   \citet{Bresolin2004ApJ615}     \\ 
  5 &  -1.057 &  -0.201 &   0.124 &  -0.337 &   8.67 &   8.14 &   0.44 &  2 &    M51 CCM 72            &   \citet{Bresolin2004ApJ615}     \\ 
  6 &   0.053 &   0.097 &   0.282 &  -0.469 &   8.53 &   8.10 &   0.66 &  2 &    M51 CCM 84A           &   \citet{Bresolin2004ApJ615}     \\ 
  7 &   0.727 &   0.316 &  -0.534 &  -0.638 &   8.23 &   7.03 &   1.08 &  3 &    NGC1232  no 04        &   \citet{Bresolin2005AA441}      \\
\hline
\end{tabular}\\
\end{center}
\end{table*}

\setcounter{table}{0}
\begin{table*}
\caption[]{Continued}
\begin{center}
\begin{tabular}{rrrrrccccll} \hline \hline
n                     &
logR$_3$              &
logR$_2$              &
logN$_2$              &
logS$_2$              &
12+log(O/H)           &
12+log(N/H)           &
t$_3$                 &
j$_T$                 &
H\,{\sc ii} region            &
reference         \\ \hline
 59 &   1.026 &   0.048 &  -0.898 &  -0.793 &   8.15 &   6.95 &   1.32 &  1 &    Tol 1008-286          &   \citet{Campbell1986MNRAS223}   \\ 
 60 &   0.313 &   0.471 &  -0.181 &  -0.450 &   8.46 &   7.33 &   0.83 &  2 &    NGC 628 H13           &   \citet{Castellan2002MNRAS329}  \\ 
 61 &   0.049 &   0.502 &   0.074 &  -0.218 &   8.50 &   7.55 &   0.78 &  2 &    NGC 1232 CDT3         &   \citet{Castellan2002MNRAS329}  \\ 
 62 &   0.647 &   0.411 &  -0.562 &  -0.264 &   8.14 &   6.85 &   1.15 &  1 &    U 5005 (2-2)          &   \citet{deBlok1998AA335}        \\ 
 63 &   0.604 &   0.013 &  -0.414 &  -0.674 &   8.31 &   7.45 &   0.91 &  2 &    M101 NGC5461          &   \citet{Esteban2009ApJ700}      \\ 
 64 &   0.111 &   0.130 &  -0.105 &  -0.587 &   8.46 &   7.65 &   0.70 &  2 &    M101 H1013            &   \citet{Esteban2009ApJ700}      \\ 
 65 &   0.576 &   0.155 &  -0.371 &  -0.801 &   8.47 &   7.46 &   0.83 &  1 &    M31  K9323            &   \citet{Esteban2009ApJ700}      \\
\hline
\end{tabular}\\
\end{center}
\end{table*}

\setcounter{table}{0}
\begin{table*}
\caption[]{Continued}
\begin{center}
\begin{tabular}{rrrrrccccll} \hline \hline
n                     &
logR$_3$              &
logR$_2$              &
logN$_2$              &
logS$_2$              &
12+log(O/H)           &
12+log(N/H)           &
t$_3$                 &
j$_T$                 &
H\,{\sc ii} region            &
reference         \\ \hline
122 &   0.849 &   0.057 &  -1.136 &  -0.710 &   7.91 &   6.50 &   1.43 &  3 &    TOL 1924-416 no 2     &   \citet{Guseva2011AA529}        \\ 
123 &   0.890 &  -0.299 &  -1.621 &  -1.060 &   7.72 &   6.25 &   1.70 &  1 &    PHL 293 B             &   \citet{Guseva2011AA529}        \\ 
124 &   0.997 &  -0.321 &  -1.358 &  -1.036 &   7.93 &   6.70 &   1.52 &  1 &    TOL 0513-393          &   \citet{Guseva2011AA529}        \\ 
125 &   0.614 &   0.338 &  -0.550 &  -0.344 &   8.11 &   6.90 &   1.13 &  1 &    J154054+565138        &   \citet{Hagele2008MNRAS383}     \\ 
126 &   0.913 &  -0.071 &  -1.243 &  -0.870 &   8.01 &   6.59 &   1.35 &  3 &    J161623+470202        &   \citet{Hagele2008MNRAS383}     \\ 
127 &   0.836 &   0.246 &  -0.536 &  -0.638 &   8.22 &   7.12 &   1.13 &  3 &    J172906+565319        &   \citet{Hagele2008MNRAS383}     \\ 
128 &   0.704 &   0.318 &  -0.418 &  -0.516 &   8.29 &   7.18 &   1.03 &  1 &    M101 NGC 5455         &   \citet{Hawley1978ApJ224}       \\
\hline
\end{tabular}\\
\end{center}
\end{table*}

\setcounter{table}{0}
\begin{table*}
\caption[]{Continued}
\begin{center}
\begin{tabular}{rrrrrccccll} \hline \hline
n                     &
logR$_3$              &
logR$_2$              &
logN$_2$              &
logS$_2$              &
12+log(O/H)           &
12+log(N/H)           &
t$_3$                 &
j$_T$                 &
H\,{\sc ii} region            &
reference         \\ \hline
185 &   0.853 &   0.176 &  -1.105 &  -0.724 &   8.06 &   6.51 &   1.29 &  1 &    Mrk 724               &   \citet{Izotov2004ApJ602}       \\ 
186 &   1.029 &   0.024 &  -1.233 &  -0.854 &   8.08 &   6.59 &   1.41 &  1 &    UM 439                &   \citet{Izotov2004ApJ602}       \\ 
187 &   0.726 &   0.442 &  -0.851 &  -0.487 &   8.09 &   6.52 &   1.25 &  1 &    Pox 36                &   \citet{Izotov2004ApJ602}       \\ 
188 &   0.905 &   0.035 &  -1.120 &  -0.907 &   8.01 &   6.62 &   1.36 &  2 &    Mrk 1315              &   \citet{Izotov2004ApJ602}       \\ 
189 &   0.628 &   0.494 &  -0.633 &  -0.511 &   8.26 &   6.78 &   1.07 &  1 &    HS 1213+3636A         &   \citet{Izotov2004ApJ602}       \\ 
190 &   0.887 &   0.148 &  -1.113 &  -0.783 &   8.04 &   6.53 &   1.33 &  1 &    HS 1214+3801          &   \citet{Izotov2004ApJ602}       \\ 
191 &   0.864 &   0.094 &  -1.037 &  -0.836 &   8.14 &   6.71 &   1.20 &  2 &    Mrk 1329              &   \citet{Izotov2004ApJ602}       \\
\hline
\end{tabular}\\
\end{center}
\end{table*}

\setcounter{table}{0}
\begin{table*}
\caption[]{Continued}
\begin{center}
\begin{tabular}{rrrrrccccll} \hline \hline
n                     &
logR$_3$              &
logR$_2$              &
logN$_2$              &
logS$_2$              &
12+log(O/H)           &
12+log(N/H)           &
t$_3$                 &
j$_T$                 &
H\,{\sc ii} region            &
reference         \\ \hline
248 &   0.470 &   0.207 &  -1.001 &  -0.614 &   7.66 &   6.24 &   1.45 &  1 &    DDO 154 no 2          &   \citet{Kennicutt2001AJ121}     \\ 
249 &   0.137 &   0.274 &  -0.066 &  -0.541 &   8.49 &   7.60 &   0.72 &  2 &    M101 H 1013           &   \citet{Kennicutt2003ApJ591}    \\ 
250 &   0.624 &   0.267 &  -0.352 &  -0.618 &   8.38 &   7.33 &   0.92 &  2 &    M101 H 1105           &   \citet{Kennicutt2003ApJ591}    \\ 
251 &   0.625 &   0.297 &  -0.503 &  -0.526 &   8.30 &   7.10 &   0.97 &  1 &    M101 H 1159           &   \citet{Kennicutt2003ApJ591}    \\ 
252 &   0.691 &   0.204 &  -0.550 &  -0.638 &   8.28 &   7.14 &   1.00 &  1 &    M101 H 1176           &   \citet{Kennicutt2003ApJ591}    \\ 
253 &   0.799 &   0.179 &  -0.831 &  -0.793 &   8.16 &   6.83 &   1.15 &  1 &    M101 H 1216           &   \citet{Kennicutt2003ApJ591}    \\ 
254 &   0.716 &   0.161 &  -0.558 &  -0.633 &   8.26 &   7.16 &   1.02 &  2 &    M101 H 128            &   \citet{Kennicutt2003ApJ591}    \\
\hline
\end{tabular}\\
\end{center}
\end{table*}

\setcounter{table}{0}
\begin{table*}
\caption[]{Continued}
\begin{center}
\begin{tabular}{rrrrrccccll} \hline \hline
n                     &
logR$_3$              &
logR$_2$              &
logN$_2$              &
logS$_2$              &
12+log(O/H)           &
12+log(N/H)           &
t$_3$                 &
j$_T$                 &
H\,{\sc ii} region            &
reference         \\ \hline
311 &   0.707 &   0.152 &  -0.885 &  -0.914 &   8.24 &   6.83 &   1.02 &  1 &    IC 10  Fib 90         &   \citet{LopezSanchez2011MN411}  \\ 
312 &   0.726 &   0.104 &  -0.917 &  -0.924 &   8.23 &   6.84 &   1.03 &  1 &    IC 10  Fib 92         &   \citet{LopezSanchez2011MN411}  \\ 
313 &   0.954 &   0.009 &  -1.001 &  -0.721 &   8.10 &   6.83 &   1.30 &  1 &    M101 NGC 5471         &   \citet{Luridiana2002RevMex38}  \\ 
314 &   0.692 &  -0.315 &  -0.636 &  -0.775 &   8.31 &   7.56 &   0.92 &  1 &    IC 10 no 120          &   \citet{Magrini2009MNRAS398}    \\ 
315 &   0.675 &   0.250 &  -0.648 &  -0.513 &   8.07 &   6.86 &   1.17 &  1 &    NGC 2403 -133,-146    &   \citet{McCall1985ApJS57}       \\ 
316 &   0.744 &   0.396 &  -0.962 &  -0.456 &   7.95 &   6.36 &   1.40 &  1 &    KISSR 396             &   \citet{Melbourne2004AJ127}     \\ 
317 &   0.876 &   0.103 &  -1.136 &  -0.614 &   7.88 &   6.45 &   1.51 &  1 &    KISSR 675             &   \citet{Melbourne2004AJ127}     \\
\hline
\end{tabular}\\
\end{center}
\end{table*}

\setcounter{table}{0}
\begin{table*}
\caption[]{Continued}
\begin{center}
\begin{tabular}{rrrrrccccll} \hline \hline
n                     &
logR$_3$              &
logR$_2$              &
logN$_2$              &
logS$_2$              &
12+log(O/H)           &
12+log(N/H)           &
t$_3$                 &
j$_T$                 &
H\,{\sc ii} region            &
reference         \\ \hline
374 &   0.598 &   0.025 &  -1.358 &  -0.824 &   7.44 &   5.94 &   1.84 &  1 &    1415+437              &   \citet{Thuan1995ApJ445}        \\ 
375 &   0.810 &   0.352 &  -0.689 &  -0.284 &   8.00 &   6.72 &   1.38 &  1 &    1533+469              &   \citet{Thuan1995ApJ445}        \\ 
376 &   0.671 &   0.023 &  -1.308 &  -0.818 &   7.57 &   6.09 &   1.70 &  1 &    S 1415+437 1.5x5      &   \citet{Thuan1999ApJ525}        \\ 
377 &   0.689 &  -0.080 &  -1.414 &  -0.910 &   7.59 &   6.10 &   1.66 &  1 &    S 1415+437 1.5x0.6    &   \citet{Thuan1999ApJ525}        \\ 
378 &   0.604 &   0.330 &  -0.316 &  -0.648 &   8.39 &   7.31 &   0.92 &  1 &    M101 NGC 5461         &   \citet{TorresPeimb1989ApJ345}  \\ 
379 &   0.924 &   0.120 &  -0.927 &  -0.754 &   8.18 &   6.84 &   1.21 &  2 &    M101 NGC 5471         &   \citet{TorresPeimb1989ApJ345}  \\ 
380 &   0.748 &   0.274 &  -0.835 &  -0.602 &   8.13 &   6.71 &   1.17 &  2 &    NGC 55                &   \citet{Tullmann2003AA412}      \\
\hline
\end{tabular}\\
\end{center}
\end{table*}


\begin{thebibliography}{}

\bibitem [Aggarwal \& Keenan (1999)]{aggarwal1999} 
          Aggarwal K.M., Keenan F.P., 1999, ApJS, 123, 31 

\bibitem [Alloin et al.(1979)]{alloin1979} 
          Alloin D., Collin-Souffrin S., Joly M., Vigroux L., 1979, A\&A, 78, 200

\bibitem [Andrievsky et al.(2002a)]{Andrievsky2002a}           
          Andrievsky S.M., et al., 2002a, A\&A, 381, 32
          
\bibitem [Andrievsky et al.(2002b)]{Andrievsky2002b} 
          Andrievsky S.M., Bersier D., Kovtyukh V.V., Luck R.E., Maciel W.J., L\'{e}pine J.R.D., 
          Beletsky Y.V., 2002b, A\&A, 384, 140

\bibitem [Andrievsky et al.(2002c)]{Andrievsky2002c} 
          Andrievsky S.M., Kovtyukh V.V., Luck R.E., L\'{e}pine J.R.D.,  Maciel W.J., 
          Beletsky Y.V., 2002c, A\&A, 392, 491  

\bibitem [Andrievsky et al.(2004)]{Andrievsky2004}           
          Andrievsky S.M., Luck R.E., Martin P., L\'{e}pine J.R.D., 2004, A\&A, 413, 159

\bibitem [Asplund et al.(2009)]{Asplund2009}
          Asplund M., Grevesse N., Sauval A.J.,  Scott P., 2009, ARA\&A, 47, 481
	
\bibitem [Baldwin, Phillips \& Terlevich (1981)]{baldwin1981}
          Baldwin J.A., Phillips M.M., Terlevich R., 1981, PASP, 93, 5

\bibitem [Bresolin et al.(1999)]{Bresolin1999ApJ540}
          Bresolin F., Kennicutt R.C., Garnett D.R., 1999, ApJ, 510, 104

\bibitem [Bresolin \& Kennicutt (2002)]{Bresolin2002ApJ572}  
          Bresolin F., Kennicutt R.C., 2002, ApJ, 572, 838

\bibitem [Bresolin et al.(2004)]{Bresolin2004ApJ615}  
          Bresolin F., Garnett D.R., Kennicutt R.C.,  2004, ApJ, 615, 228

\bibitem [Bresolin et al.(2005)]{Bresolin2005AA441}
          Bresolin F., Schaerer D., Gonz\'{a}lez Delgado R.M., Stasi\'{n}ska G., 
          2005, A\&A, 441, 981

\bibitem [Bresolin  (2007)]{Bresolin2007ApJ656}
          Bresolin F., 2007, ApJ, 656, 186

\bibitem [Bresolin et al.(2009a)]{Bresolin2009ApJ695}  
          Bresolin F., Ryan-Weber E., Kennicutt R.C., Goddard Q., 
          2009a, ApJ, 695, 580

\bibitem [Bresolin et al.(2009b)]{Bresolin2009ApJ700}
          Bresolin F., Gieren W., Kudritzki R.-P., Pietrzy\'{n}ski G., 
          Urbaneja M.A., Carraro G., 2009b, ApJ, 700, 309

\bibitem [Campbell et al.(1986)]{Campbell1986MNRAS223}  
          Campbell A., Terlevich R., Melnick J., 1986, MNRAS, 223, 811 

\bibitem [Castellanos et al.(2002)]{Castellan2002MNRAS329}  
          Castellanos M., D\'{i}az A.I., Terlevich E., 2002, MNRAS, 329, 315

\bibitem [Cid Fernandes et al.(2011)]{CidFernandes2011MNRAS413}
           Cid Fernandes R., Stasi\'{n}ska G., Mateus A.,  Asari N.V., 
          2011, MNRAS, 413, 1687 


\bibitem [de Blok \& van der Hulst (1998)]{deBlok1998AA335}   
          de Blok W.J.G., van der Hulst J.M., 1998, A\&A, 335, 421

\bibitem [Dopita \& Evans (1986)]{Dopita1986ApJ307}
          Dopita M.A., Evans I.N., 1986, ApJ, 307, 431 

\bibitem [Dufour et al.(1980)]{Dufour1980ApJ236}
          Dufour R.J., Talbot R.J., Jensen E.B., Shields G.A., 1980, ApJ, 236, 119 

\bibitem [Edl\'{e}n (1985)]{edlen1985} 
          Edl\'{e}n B., 1985, Phys. Scripta, 31, 345 

\bibitem [Esteban et al.(2004)]{Esteban2004MN355}
          Esteban C., Peimbert M., Garc\'{\i}a-Rojas J., Ruiz M. T., Peimbert A.,  
          Rodr\'{\i}guez M., 2004, MNRAS, 355, 229

\bibitem [Esteban et al.(2009)]{Esteban2009ApJ700}
          Esteban C., Bresolin F., Peimbert M., Garc\'{i}a-Rojas J., Peimbert A., Mesa-Delgado A., 
          2009, ApJ, 700, 654

\bibitem [Ferguson et al.(1998)]{Ferguson1998AJ116}  
          Ferguson A.M.N., Gallagher J.S., Wyse R.F.G., 1998, AJ, 116, 673 

\bibitem [Fierro et al.(1986)]{Fierro1986PASP98}  
          Fierro J., Torres-Peimbert S., Peimbert M., 1986, PASP, 98, 1032

\bibitem [French (1980)]{French1980ApJ240} 
          French H.B., 1980, ApJ, 240, 41

\bibitem [Fricke et al.(2001)]{Fricke2001AJ121}
          Fricke K.J., Izotov Y.I., Papaderos P., Guseva N.G., Thuan T.X.,
          2001, AJ, 121, 169

\bibitem [Froese Fisher \& Tachiev (2004)]{froese2004}  
          Froese Fischer C., Tachiev G., 2004, ADNDT, 87, 1 

\bibitem [Henry et al.(2000)]{Henry2000ApJ541} 
          Henry R.B.C., Edmunds M.G., K\'{o}ppen J., 2000, ApJ, 541, 660

\bibitem [Galav\'{i}s et al.(1997)]{galavis1997} 
          Galav\'{i}s M.E., Mendoza C., Zeippen C.J., 1997, A\&AS, 123, 159  


\bibitem [Garc\'{\i}a-Rojas et al.(2004)]{GarciaRojas2004ApJS153}
         Garc\'{\i}a-Rojas J., Esteban C., Peimbert M., Rodr\'{\i}guez M., 
         Ruiz M. T., Peimbert A. 2004, ApJS, 153, 501

\bibitem [Garc\'{\i}a-Rojas et al.(2005)]{GarciaRojas2005MN362}
         Garc\'{\i}a-Rojas J., Esteban C., Peimbert A., Peimbert M., Rodr\'{\i}guez M., Ruiz M.T., 
         2005, MNRAS, 362, 301

\bibitem [Garc\'{\i}a-Rojas et al.(2006)]{GarciaRojas2006MN368}
          Garc\'{\i}a-Rojas J., Esteban C., Peimbert M., Costado M.T., Rodr\'{\i}guez M., Peimbert A.,
          Ruiz M. T. 2006, MNRAS, 368, 253

\bibitem [Garc\'{\i}a-Rojas \& Esteban (2007)]{GarciaRojas2007ApJ670}
          Garc\'{\i}a-Rojas J., Esteban C., 2007, ApJ, 670, 457

\bibitem [Garnett (1992)]{garnett1992} 
          Garnett D.R., 1992, AJ, 103, 1330

\bibitem [Garnett et al.(1997)]{Garnett1997ApJ489}  
          Garnett D.R., Shields G.A., Skillman E.D., Sagan S.P., Dufour R.J., 
          1997, ApJ, 489, 63

\bibitem [Garnett et al.(2004)]{Garnett2004ApJ607}  
          Garnett D.R., Kennicutt R.C., Bresolin F., 2004, ApJ, 607, L21

\bibitem [Goddard et al.(2011)]{Goddard2011MNRAS412}  
          Goddard Q., Bresolin F., Kennicutt R.C., Ryan-Weber E.V., Rosales-Ortega F.F., 
          2011, MNRAS, 412, 1246

\bibitem [Gonz\'{a}lez-Delgado  et al.(1994)]{ConzalezDelgado1994ApJ437}
          Gonz\'{a}lez-Delgado  R.M., et al., 1994, ApJ, 437, 239

\bibitem [Guseva et al.(2000)]{Guseva2000ApJ531}
          Guseva N.G., Izotov Y.I., Thuan T.X., 2000, ApJ, 531, 776 

\bibitem [Guseva et al.(2001)]{Guseva2001AA378}
          Guseva N.G., et al., 2001, A\&A, 378, 756

\bibitem [Guseva et al.(2003a)]{Guseva2003AA407p91} 
          Guseva N.G., Papaderos P., Izotov Y.I., Green R.F., Fricke K.J., Thuan T.X., Noeske K.G.,
          2003a, A\&A, 407, 91 

\bibitem [Guseva et al.(2003b)]{Guseva2003AA407p105}
          Guseva N.G., Papaderos P., Izotov Y.I., Green R.F., Fricke K.J., Thuan T.X., Noeske K.G.,
          2003b, A\&A, 407, 105 

\bibitem [Guseva et al.(2004)]{Guseva2004AA421} 
          Guseva N.G., Papaderos P., Izotov Y.I., Noeske K.G., Fricke K.J.,
          2004, A\&A, 421, 519

\bibitem [Guseva et al.(2009)]{Guseva2009AA505} 
          Guseva N.G., Papaderos P., Meyer H.T., Izotov Y.I., Fricke K.J., 
          2009, A\&A, 505, 63 

\bibitem [Guseva et al.(2011)]{Guseva2011AA529}
          Guseva N.G., Izotov Y.I., Stasi\'{n}ska G., Fricke K.J., Henkel C., Papaderos P., 
          2011, A\&A, 529, A149

\bibitem [H\"{a}gele et al.(2008)]{Hagele2008MNRAS383}
          H\"{a}gele G.F., D\'{i}az \'{A}.I., Terlevich E., Terlevich R., P\'{e}rez-Montero E., Cardaci M.V.,
          2008, MNRAS, 383, 209

\bibitem [Hawley (1978)]{Hawley1978ApJ224} 
          Hawley S.A., 1978, ApJ, 224, 417

\bibitem [Hodge \& Miller (1995)]{Hodge1995ApJ451}
          Hodge P., Miller B.W., 1995, ApJ, 451, 176

\bibitem [Hudson \& Bell (2005)]{hudson2005} 
          Hudson C.E., Bell K.L., 2005,  A\&A,  430, 725 

\bibitem [Izotov et al.(1994)]{Izotov1994ApJ435}
          Izotov Y.I., Thuan T.X., Lipovetsky V.A., 1994, ApJ, 435, 647

\bibitem [Izotov et al.(1997)]{Izotov1997ApJS108}
          Izotov Y.I., Thuan T.X., Lipovetsky V.A., 1997, ApJS, 108, 1

\bibitem [Izotov \& Thuan (1998a)]{Izotov1998ApJ497}
          Izotov Y.I., Thuan T.X., 1998a, ApJ, 497, 227

\bibitem [Izotov \& Thuan (1998b)]{Izotov1998ApJ500}
          Izotov Y.I., Thuan T.X., 1998b, ApJ, 500, 188

\bibitem [Izotov et al.(1999)]{Izotov1999ApJ527}
          Izotov Y.I., Chaffee F.H., Foltz C.B., Green R.F., Guseva N.G., Thuan T.X., 1999, ApJ, 527, 757

\bibitem [Izotov et al.(2001)]{Izotov2001ApJ562}
          Izotov Y.I., Chaffee F.H., Green R.F., 2001, ApJ, 562, 727

\bibitem [Izotov et al.(2004)]{Izotov2004AA421} 
          Izotov Y.I., Papaderos P., Guseva N. G., Fricke K.J., Thuan T.X., 
          2004, A\&A, 421, 539

\bibitem [Izotov \& Thuan (2004)]{Izotov2004ApJ602}
          Izotov Y.I., Thuan T.X., 2004, ApJ, 602, 200

\bibitem [Izotov et al.(2005)]{Izotov2005ApJ632}
          Izotov Y.I., Thuan T.X.,  Guseva N.G., 2005, ApJ, 632, 210

\bibitem [Izotov et al.(2006)]{Izotov2006AA448}
          Izotov Y.I., Stasi\'{n}ska G., Meynet G., Guseva N.G.,   
          Thuan T.X., 2006, A\&A, 448, 955

\bibitem [Izotov et al.(2009)]{Izotov2009AA503} 
          Izotov Y.I., Guseva N.G., Fricke K.J.,  Papaderos P., 
          2009, A\&A, 503, 61 

\bibitem [Izotov et al.(2011)]{Izotov2011AA533} 
          Izotov Y.I., Guseva N.G., Fricke K.J.,  Henkel C., 
          2011, A\&A, 533, A25 

\bibitem [Johansson et al.(1992)]{johansson1992} 
          Johansson L., Magnusson C.E., Joelsson I., Zetterberg P.O., 
          1992,  Phys. Scripta, 46, 221

\bibitem [Kauffmann et al.(2003)]{kauffmann2003}
          Kauffmann G., et al., 2003, MNRAS, 346, 1055

\bibitem [Kehrig et al.(2004)]{Kehrig2004AJ128} 
          Kehrig C., Telles E., Cuisinier F., 2004, AJ, 128, 1141

\bibitem [Kehrig et al.(2011)]{Kehrig2011AA526} 
          Kehrig C., et al., 2011, A\&A, 526, A128

\bibitem [Kennicutt \& Skillman (2001)]{Kennicutt2001AJ121}
          Kennicutt R.C., Skillman E.D., 2001, AJ, 121, 1461 

\bibitem [Kennicutt et al.(2003)]{Kennicutt2003ApJ591}
          Kennicutt R.C., Bresolin F., Garnett D.R., 2003, ApJ, 591, 801

\bibitem [Kewley et al.(2001)]{kewley2001}
          Kewley L.J., Dopita M.A., Sutherland R.S., Heisler C.A., Trevena J.,  
          2001, ApJ, 556, 121

\bibitem [Kewley \& Dopita (2002)]{Kewley2002ApJS142}
          Kewley L.J., Dopita M.A., 2002, ApJS, 142, 35

\bibitem [Kewley \& Ellison (2008)]{Kewley2008ApJ681}
          Kewley L.J., Ellison S.L., 200, ApJ, 681, 1183 

\bibitem [Kinkel \& Rosa (1994)]{Kinkel1994AA282} 
          Kinkel U., Rosa M.R., 1994, A\&A, 282, L37

\bibitem [Kniazev et al.(2000)]{Kniazev2000AA357} 
          Kniazev A.Y., et al., 2000, A\&A, 357, 101

\bibitem [Kniazev et al.(2004)]{Kniazev2004ApJS153}
          Kniazev A.Y., Pustilnik S.A., Grebel E.K., Lee H., 
          Pramskij A.G., 2004, ApJS, 153, 429 

\bibitem [Kobulnicky et al.(1997)]{Kobulnicky1997ApJ477}
          Kobulnicky H.A., Skillman E.D., Roy J.-R., Walsh J.R., Rosa M.R., 
          1997, ApJ, 477, 679

\bibitem [Kobulnicky \& Skillman (1998)]{Kobulnicky1998ApJ497}
          Kobulnicky H.A., Skillman E.D., 1998, ApJ, 497, 601

\bibitem [Kunth \& Sargent (1983)]{Kunth1983ApJ273}
          Kunth D., Sargent W.L.W., 1983, ApJ, 273, 81

\bibitem [Kwitter \& Aller (1981)]{Kwitter1981MNRAS195} 
          Kwitter K.B., Aller L.H., 1981, MNRAS, 195, 939

\bibitem [Lee et al.(2003a)]{LeeHenry2003AJ125p2975} 
          Lee H.,  McCall M.L., Richer M.G., 2003a, AJ, 125, 2975

\bibitem [Lee et al.(2003b)]{LeeHenry2003AA401} 
          Lee H.,  Grebel E.K., Hodge P.W., 2003b, A\&A, 401, 141 

\bibitem [Lee et al.(2005)]{LeeHenry2005ApJ620} 
          Lee H., Skillman E.D., Venn K.A., 2005, ApJ, 620, 223

\bibitem [Lee et al.(2004)]{LeeJanice2004ApJ616} 
          Lee J.C., Salzer J.J., Melbourne J., 2004, ApJ, 616, 752

\bibitem [Lequeux et al.(1979)]{Lequeux1979AA80} 
          Lequeux J., Peimbert M., Rayo J.F., Serrano A., Torres-Peimbert S., 
          1979, A\&A, 80, 155  

\bibitem [L\'{o}pez-S\'{a}nchez et al.(2004)]{LopezSanchez2004ApJS153} 
          L\'{o}pez-S\'{a}nchez \'{A}.R.,  Esteban C., Rodr\'{i}guez M., 2004, ApJS, 153, 243 

\bibitem [L\'{o}pez-S\'{a}nchez et al.(2007)]{LopezSanchez2007ApJ656} 
          L\'{o}pez-S\'{a}nchez \'{A}.R.,  Esteban C., Garc\'{i}a-Rojas J., 
          Peimbert M., Rodr\'{i}guez M., 2007, ApJ, 656, 168 

\bibitem [L\'{o}pez-S\'{a}nchez \& Esteban(2009)]{LopezSanchez2009AA508}
          L\'{o}pez-S\'{a}nchez \'{A}.R.,  Esteban C., 2008, A\&A, 508, 615

\bibitem [L\'{o}pez-S\'{a}nchez \& Esteban(2010)]{Lopezsanchez2010AA517}
          L\'{o}pez-S\'{a}nchez \'{A}.R.,  Esteban C., 2010, A\&A, 517, A85

\bibitem [L\'{o}pez-S\'{a}nchez et al.(2011)]{LopezSanchez2011MN411} 
          L\'{o}pez-S\'{a}nchez \'{A}.R.,  Mesa-Delgado A., L\'{o}pez-Mart\'{i}n L., Esteban C., 
          2011, MNRAS, 411, 2076 

\bibitem [Luridiana et al.(2002)]{Luridiana2002RevMex38}
          Luridiana V., Esteban C., Peimbert M., Peimbert A., 2002,
          Rev. Mex. A.A, 38, 97 

\bibitem [Maeder (1992)]{Maeder1992AA264}  
          Maeder A., 1992, A\&A, 264, 105

\bibitem [Magrini \& Goncalves (2009)]{Magrini2009MNRAS398} 
          Magrini L., Goncalves D.R., 2009, MNRAS, 398, 280
          
\bibitem [Mattsson (2010)]{Mattsson2010} 
          Mattsson L., 2010, A\&A, 515, A68

\bibitem [McCall et al.(1985)]{McCall1985ApJS57}
          McCall M.L., Rybski P.M., Shields G.A., 1985, ApJS, 57, 1

\bibitem [McGaugh (1991)]{McGaugh1991ApJ380}
          McGaugh S.S., 1991, ApJ, 380,, 140

\bibitem [Melbourne et al.(2004)]{Melbourne2004AJ127}
          Melbourne J., Phillips A., Salzer J.J., Gronwall C., Sarajedini V.L., 2004, AJ, 127, 686

\bibitem [Melnick et al.(1992)]{Melnick1992AA253} 
          Melnick J., Heydari-Malayeri M., Leisy P., 1992, A\&A, 253, 16

\bibitem [Mendoza \& Zeippen (1982)]{mendoza1982}
          Mendoza C., Zeippen C.J., 1982, MNRAS, 199, 1025

\bibitem [Miller (1996)]{Miller1996AJ112} 
          Miller B.W., 1996, AJ, 112, 991

\bibitem [Moustakas et al.(2010)]{Moustakas2010ApJS190} 
          Moustakas J., Kennicutt R.C., Tremonti C.A., Dale D.A., 
          Smith J.-D.T., Calzetti D., 2010. ApJS, 190, 233 

\bibitem [Noeske et al.(2000)]{Noeske2000AA361} 
          Noeske K.G., Guseva N.G.,  Fricke K.J., Izotov Y.I.,  Papaderos P., Thuan T.X., 
          2000, A\&A, 361, 33 

\bibitem [Pagel (1997)]{Pagel1997book} 
          Pagel B.E.J., 1997, Nucleosynthesis and Chemical Evolution of Galaxies
         (Cambridge: Cambridge Univ. Press)

\bibitem [Pagel et al.(1979)]{Pagel1979MNRAS189} 
          Pagel B.E.J., Edmunds M.G., Blackwell D.E., Chun M.S., Smith G., 1979, MNRAS, 189, 95

\bibitem [Pagel et al.(1980)]{Pagel1980MNRAS193}
          Pagel B.E.J., Edmunds M.G., Smith G., 1980, MNRAS, 193, 219

\bibitem [Pagel et al.(1992)]{Pagel1992MNRAS255} 
          Pagel B.E.J., Simonson E.A., Terlevich R.J., Edmunds M.G., 1992, MNRAS, 255, 325

\bibitem [Pastoriza et al.(1993)]{Pastoriza1993MNRAS260}
          Pastoriza M.G., Dottori H., Terlevich E., Terlevich R., D\'{i}az A.I. 
          1993, MNRAS, 260, 177

\bibitem [Peimbert (2003)]{Peimbert2003ApJ584} 
          Peimbert A. 2003, ApJ, 584, 735

\bibitem [Peimbert et al.(2005)]{Peimbert2005ApJ634} 
          Peimbert A., Peimbert M., Ruiz M.T. 2005, ApJ, 634, 1056

\bibitem [Peimbert (1967)]{Peimbert1967ApJ150}
          Peimbert M., 1967, ApJ, 150, 825 

\bibitem [Peimbert \& Costero (1969)]{PeimbertCostero1969}
          Peimbert M., Costero R., 1969, Bol. Obs. Tonantzintla y Tacubaya, 5, 3

\bibitem [Peimbert et al.(1986)]{Peimbert1986AA158} 
          Peimbert M., Pena M., Torres-Peimbert S., 1986, A\&A, 158, 266

\bibitem [Pe\~{n}a et al.(2007)]{Pena2007AA476} 
          Pe\~{n}a M., Stasi\'{n}ska G., Richer M.G., 2007, A\&A, 476, 745

\bibitem [Pe\~{n}a-Guerrero et al.(2012)]{PenaGuerrero2012ApJ746} 
          Pe\~{n}a-Guerrero M.A., Peimbert A., Peimbert M., Ruiz M.T. 
          2012, ApJ, 746, 115

\bibitem [P\'{e}rez-Montero et al.(2009)]{PerezMontero2009AA497}
          P\'{e}rez-Montero E., Garc\'{i}a-Benito R., D\'{i}az A.I., P\'{e}rez E., Kehrig C., 
          2009, A\&A, 497, 53
 
\bibitem [Pettini \& Pagel (2004)]{Pettini2004MNRAS348}
          Pettini M., Pagel B.E.J., 2004, MNRAS, 348, 59L

\bibitem [Pilyugin (2000)]{Pilyugin2000AA362}
          Pilyugin L.S., 2000, A\&A, 362, 325 
		  
\bibitem [Pilyugin (2001)]{Pilyugin2001AA369}
          Pilyugin L.S., 2001, A\&A, 369, 594

\bibitem [Pilyugin (2003a)]{Pilyugin2003AA397a} 
          Pilyugin L.S., 2003, A\&A, 397, 109

\bibitem [Pilyugin (2003b)]{Pilyugin2003AA399} 
          Pilyugin L.S., 2003, A\&A, 399, 1003 

\bibitem [Pilyugin et al.(2003)]{Pilyugin2003AA397} 
          Pilyugin L.S., Thuan, T.X., V\'{\i}lchez J.M., 2003, A\&A, 397, 487

\bibitem [Pilyugin et al.(2004)]{Pilyugin2004AA425} 
          Pilyugin L.S., V\'{\i}lchez J.M., Contini T., 2004, A\&A, 425, 849 

\bibitem [Pilyugin \& Thuan (2005)]{Pilyugin2005ApJ631} 
          Pilyugin L.S., Thuan T.X., 2005, ApJ, 631, 231

\bibitem [Pilyugin et al.(2006)]{Pilyugin2006MNRAS367} 
          Pilyugin L.S.,Thuan T.X., V\'{\i}lchez J.M., 2006, MNRAS, 367, 1139 

\bibitem [Pilyugin et al.(2010)]{pilyugin2010ApJ720} 
          Pilyugin L.S., V\'{i}lchez J.M., Thuan T.X., 2010, ApJ, 720, 1738 

\bibitem [Pilyugin \& Mattsson (2011)]{Pilyugin2011MNRAS412} 
          Pilyugin L.S., Mattsson L., 2011, MNRAS, 412, 1145

\bibitem [Pilyugin \& Thuan (2011)]{Pilyugin2011ApJ726} 
          Pilyugin L.S., Thuan T.X., 2011, ApJ, 726, L23

\bibitem [Pilyugin et al.(2012)]{Pilyugin2012MNRAS000} 
          Pilyugin L.S., V\'{\i}lchez J.M., Mattsson L., Thuan T.X., 2012, MNRAS, 421, 1624 

\bibitem [Popescu \& Hopp (2000)]{Popescu2000AAS142}
          Popescu C.C., Hopp U., 2000, A\&AS, 142, 247

\bibitem [Pradhan et al.(2006)]{pradhan2006}  
          Pradhan A.K., Montenegro M., Nahar S.N., Eissner, W.,  
          2006, MNRAS, 366, L6 

\bibitem [Pustilnik et al.(2002)]{Pustilnik2002AA389}
          Pustilnik S.A., Kniazev A.Y., Masegosa J., M\'{a}rquez I M., Pramskij A.G., Ugryumov A.V., 
          2002, A\&A, 389, 779

\bibitem [Pustilnik et al.(2003a)]{Pustilnik2003AA400}
          Pustilnik S., Zasov A., Kniazev A., Pramskij A., Ugryumov A., Burenkov A., 
          2003a, A\&A, 400, 841

\bibitem [Pustilnik et al.(2003b)]{Pustilnik2003AA409}
          Pustilnik S.A., Kniazev A.Y., Pramskij A.G., Ugryumov A.V., Masegosa J., 
          2003b, A\&A, 409, 917

\bibitem [Pustilnik et al.(2005)]{Pustilnik2005AA443}
          Pustilnik S.A., Kniazev A.Y., Pramskij A.G., 2005, A\&A, 443, 91

\bibitem [Pustilnik et al.(2006)]{Pustilnik2006AstL32}
          Pustilnik S.A., Engels D., Kniazev A.Y., Pramskij A.G., Ugryumov A.V., Hagen H.-J., 
          2006, AstL, 32, 228

\bibitem [Rayo et al.(1982)]{Rayo1982ApJ255} 
          Rayo J.F., Peimbert M., Torres-Peimbert S., 1982, ApJ, 255, 1

\bibitem [Rodr\'{i}guez \& Garc\'{i}a-Rojas (2010)]{Rodriguez2010ApJ708} 
          Rodr\'{i}guez M., Garc\'{i}a-Rojas J., 2010, ApJ, 708, 1551

\bibitem [Saviane et al.(2008)]{Saviane2008AA487} 
          Saviane I., Ivanov V.D., Held E.V., Alloin D., Rich R.M., Bresolin F., Rizzi L.,
          2008, A\&A, 487, 901

\bibitem [Sedwick \& Aller (1981)] {Sedwick1981PNAS78}
          Sedwick K.E., Aller L.H., 1981, Proc. Nat. Acad. Sci. USA., 78, 1994

\bibitem [Skillman (1985)]{Skillman1985ApJ290}
          Skillman E.D., 1985, ApJ, 290, 449

\bibitem [Skillman \& Kennicutt (1993)] {Skillman1993ApJ655}
          Skillman E.D., Kennicutt R.C., 1993, ApJ, 411, 655

\bibitem [Skillman et al.(2003)] {Skillman2003AJ125} 
          Skillman E.D., C\^{o}t\'{e} S., Miller B.W., 2003, AJ, 125, 610

\bibitem [Stanghellini et al.(2010)]{Stanghellini2010AA521}
          Stanghellini L., Magrini L., Villaver E., Galli D., 2010, A\&A, 521, A3

\bibitem [Stasi\'{n}ska (2005)]{Stasinska2005AA434} 
          Stasi\'{n}ska G.,  2005, A\&A, 434, 507 

\bibitem [Stasi\'{n}ska (2006)]{Stasinska2006AA454}
          Stasi\'{n}ska G.,  2006, A\&A, 454, L127 

\bibitem [Stasi\'{n}ska et al.(2006)]{Stasinska2006MNRAS371}
          Stasi\'{n}ska G., Cid Fernandes R., Mateus A., Sodr\'{e} L., Asari N.V.,  
          2006, MNRAS, 371, 972

\bibitem [Stasi\'{n}ska et al.(2008)]{Stasinska2008MNRAS391}
          Stasi\'{n}ska G., Asari N.V., Cid Fernandes R., Gomes J.M., Schlickmann M., 
          Mateus A., Schoenell W., Sodr\'{e} L., 2008, MNRAS, 391, L29 

\bibitem [Storey \&  Zeippen (2000)]{storey2000}
          Storey P.J., Zeippen C.J., 2000, MNRAS, 312, 813

\bibitem [Tayal \& Gupta (1999)]{tayal1999} 
          Tayal S.S., Gupta G.P., 1999, ApJ, 526, 544 

\bibitem [Terlevich et al.(1991)]{Terlevich1991AAS91}
          Terlevich R., Melnick J., Masegosa J., Moles M., Copetti M.V.F., 
          1991, A\&AS, 91, 285

\bibitem [Thuan et al.(1995)]{Thuan1995ApJ445}
          Thuan T.X., Izotov Y.I., Lipovetsky V.A., 1995, ApJ, 445, 108

\bibitem [Thuan et al.(1999)]{Thuan1999ApJ525}
          Thuan T.X., Izotov Y.I., Foltz C.B., 1999, ApJ, 525, 105 

\bibitem [Thuan et al.(2010)]{Thuan2010ApJ712} 
          Thuan T.X., Pilyugin L.S., Zinchenko I.A., 2010, ApJ , 712, 1029 

\bibitem [Torres-Peimbert et al.(1989)]{TorresPeimb1989ApJ345}
          Torres-Peimbert S., Peimbert M., Fierro J., 1989, ApJ, 345, 186

\bibitem [Tremonti et al.(2004)]{Tremonti2004ApJ613} 
          Tremonti C.A.,  et al.,  2004, ApJ, 613, 898

\bibitem [Tsamis et al. (2003)]{Tsamis2003MN338} 
          Tsamis Y.G., Balrlow M.J., Liu X.-W., Danziger I.J., Storey, P.J., 
          2003, MNRAS, 338, 687

\bibitem [T\"{u}llmann et al.(2003)]{Tullmann2003AA412} 
          T\"{u}llmann R., Rosa M.R., Elwert T., Bomans D.J., Ferguson A.M.N., Dettmar R.-J., 
          2003, A\&A, 412, 69

\bibitem [van den Hoek \& Groenewegen (1997)]{vandenhoek1997AAS123} 
          van den Hoek L.B., Groenewegen M.A.T., 1997, A\&AS, 123, 305

\bibitem [van Zee et al.(1997)]{vanZee1997AJ114} 
          van Zee L., Haynes M.P., Salzer J.J., 1997, AJ, 114, 2479 

\bibitem [van Zee et al.(1998)]{vanZee1998AJ116} 
          van Zee L., Salzer J.J., Haynes M.P., O`Donoghiu A.A., 
          Balonek T.J.,  1998, AJ, 116, 2805

\bibitem [van Zee (2000)]{vanZee2000ApJ543}
          van Zee L., 2000, ApJ, 543, L31

\bibitem [van Zee \& Haynes (2006)]{vanZee2006ApJ636} 
          van Zee L., Haynes M.P., 2006, ApJ, 636, 214

\bibitem [van Zee et al.(2006)]{vanZee2006ApJ637}
          van Zee L., Skillman E.D., Haynes M.P., 2006, ApJ, 637, 269

\bibitem [Vila-Costas \& Edmunds (1992)]{Vila1992MNRAS259} 
          Vila-Costas M.B., Edmunds M.G., 1992, MNRAS, 259, 121

\bibitem[V\'{\i}lchez et al.(1988)]{Vilchez1988MNRAS235}
         V\'{\i}lchez J.M., Pagel B.E.J., D\'{\i}az A.I., Terlevich E., 
         Edmunds M.G., 1988, MNRAS, 235, 633

\bibitem[V\'{\i}lchez et al.(2003)]{Vilchez2003ApJS145} 
         V\'{\i}lchez J.M., Iglesias-P\'{a}ramo J., 2003, ApJS, 145, 225

\bibitem [Webster \& Smith (1983)]{Webster1983MNRAS204} 
          Webster B.L., Smith M.G., 1983, MNRAS, 204, 743

\bibitem [Wen\aa ker (1990)]{wenaker1990} 
          Wen\aa ker I., 1990, Phys. Scripta, 42, 667  

\bibitem [Williams et al.(2008)]{Williams2008ApJ677} 
          Williams R., Jenkins E.B., Baldwin J.A., Zhang Y., Sharpee B., 
          Pellegrini E., Phillips M., 2008, ApJ, 677, 1100 

\bibitem [Yin et al.(2007)]{Yin2007AA462} 
          Yin S.Y., Liang Y.C., Hammer F., Brinchmann J., Zhang B.,
          Deng L.C., Flores H., 2007, A\&A, 462, 535 

\bibitem [York et al.(2000)]{York2000AJ120} 
          York D.G., et al., 2000, AJ, 120, 1579 

\bibitem [Zahid \& Bresolin (2011)]{Zahid2011ApJ141}
          Zahid H.J., Bresolin F., 2011, AJ, 141, 192

\bibitem [Zaritsky (1992)]{Zaritsky1992ApJ390} 
          Zaritsky D., 1992, ApJ, 390, L73

\bibitem [Zaritsky et al.(1994)]{Zaritsky1994ApJ420} 
          Zaritsky D., Kennicutt R.C., Huchra J.P., 1994, ApJ, 420, 87 


\end{thebibliography}
\end{document}